\documentclass[10pt,journal,final,twoside,compsoc]{IEEEtran}

\ifCLASSOPTIONcompsoc
  \usepackage[nocompress]{cite}
\else
  \usepackage{cite}
\fi
\usepackage{booktabs}       
\usepackage{hyperref}
\usepackage{url}
\usepackage{xcolor}
\usepackage{subcaption,graphicx}
\usepackage{multirow}
\usepackage{array}
\usepackage{tikz}
\usepackage{lipsum,adjustbox}
\usepackage[linesnumbered,ruled]{algorithm2e}
\usepackage{wrapfig}
\usepackage{amsmath,amssymb}
\usepackage[misc,geometry]{ifsym}

\begin{document}
\title{Pangu-Weather: A 3D High-Resolution System for Fast and Accurate Global Weather Forecast}

\author{
Kaifeng Bi,
Lingxi~Xie,
Hengheng Zhang,
Xin Chen,
Xiaotao Gu,
and Qi~Tian\textsuperscript{\Letter},~\IEEEmembership{Fellow,~IEEE}

\IEEEcompsocitemizethanks{
\IEEEcompsocthanksitem All authors are with Huawei Cloud Computing, Shenzhen, Guangdong 518129, China. \protect \\
E-mail: \{bikaifeng1,tian.qi1\}@huawei.com, 198808xc@gmail.com
\IEEEcompsocthanksitem Qi Tian is the corresponding author.
}

}

%
%

\markboth{Technical Report}%
{Bi \MakeLowercase{\textit{et al.}}, Pangu-Weather: A 3D High-Resolution Model for Fast and Accurate Global Weather Forecast}
%



\IEEEtitleabstractindextext{%
\begin{abstract}
In this paper, we present Pangu-Weather, a deep learning based system for fast and accurate global weather forecast. For this purpose, we establish a data-driven environment by downloading $43$ years of hourly global weather data from the 5th generation of ECMWF reanalysis (ERA5) data and train a few deep neural networks with about $256$ million parameters in total. The spatial resolution of forecast is $0.25^\circ\times0.25^\circ$, comparable to the ECMWF Integrated Forecast Systems (IFS). More importantly, for the first time, an AI-based method outperforms state-of-the-art numerical weather prediction (NWP) methods in terms of accuracy (latitude-weighted RMSE and ACC) of all factors (\textit{e.g.}, geopotential, specific humidity, wind speed, temperature, \textit{etc.}) and in all time ranges (from one hour to one week). There are two key strategies to improve the prediction accuracy: (i) designing a 3D Earth Specific Transformer (3DEST) architecture that formulates the height (pressure level) information into cubic data, and (ii) applying a hierarchical temporal aggregation algorithm to alleviate cumulative forecast errors. In deterministic forecast, Pangu-Weather shows great advantages for short to medium-range forecast (\textit{i.e.}, forecast time ranges from one hour to one week). Pangu-Weather supports a wide range of downstream forecast scenarios, including extreme weather forecast (\textit{e.g.}, tropical cyclone tracking) and large-member ensemble forecast in real-time. Pangu-Weather not only ends the debate on whether AI-based methods can surpass conventional NWP methods, but also reveals novel directions for improving deep learning weather forecast systems.
\end{abstract}

}

\maketitle

\IEEEdisplaynontitleabstractindextext

%

\IEEEraisesectionheading{\section{Introduction}
\label{Introduction}}
Weather forecast is one of the most important scenarios of scientific computing. It offers the ability of predicting future weather changes, especially the occurrence of extreme weather events (\textit{e.g.}, floods, droughts, hurricanes, \textit{etc.}), which has large values to the society (\textit{e.g.}, daily activity, agriculture, energy production, transportation, industry, \textit{etc.}). In the past decade, with the bloom of high-performance computational device, the community has witnessed a rapid development in the research field of numerical weather prediction (NWP)~\cite{nwp_evolution}. Conventional NWP methods mostly follow a simulation-based paradigm which formulates the physical rules of atmospheric states into partial differentiable equations (PDEs) and solves them using numerical simulations~\cite{wrf_model,ecmwf_method,ecmwf_lagrange}. Due to the high complexity of solving PDEs, these NWP methods are often very slow, \textit{e.g.}, with a spatial resolution of $0.25^{\circ}\times0.25^{\circ}$, a single simulation procedure for $10$-day forecast can take hours of computation using hundreds of nodes in a supercomputer~\cite{ecmwf_speed}. This largely reduces the timeliness in daily weather forecast and the number of ensemble members that can be used for probabilistic weather forecast. In addition, conventional NWP algorithms largely rely on the parametric numerical models, but these models, albeit being very complex~\cite{nwp_evolution}, are often considered inadequate~\cite{model_uncertainty_review,model_error_ecmwf_early}, \textit{e.g.}, errors will be introduced by parameterization of unresolved processes.

To address the above issues, a promising direction lies in data-driven weather forecast with AI, in particular, deep learning\footnote{Throughout this paper, we will use `conventional NWP' or simply `NWP' to refer to the numerical simulation methods, and use `AI-based' or `deep learning based' to specify data-driven forecast systems. We understand that, verbally, AI-based methods also belong to NWP, but we follow the convention~\cite{can_ai_beat_nwp} to use these terms.}. The methodology is to use a deep neural network to capture the relationship between the input (observed data) and output (target data to be predicted). On specialized computational device (\textit{e.g.}, GPUs), AI-based methods run very fast and easily achieve a tradeoff between model complexity, prediction resolution, and prediction accuracy~\cite{ai_weather_old_1,weatherbench,weyn_old,weyn_new,graph_weather,fourcastnet,swinvrnn}. As a recent example, FourCastNet~\cite{fourcastnet} increased the spatial resolution to $0.25^\circ\times0.25^\circ$, comparable to the ECMWF Integrated Forecast Systems (IFS), yet it takes only $7$ seconds on four GPUs for making a $100$-member, $24$-hour forecast, which is orders of magnitudes faster than the conventional NWP methods. However, the forecast accuracy of FourCastNet is still below satisfaction, \textit{e.g.}, the RMSE of $5$-day Z500 forecast using a single model and a $100$-member ensemble are $484.5$ and $462.5$, respectively, which are much worse than $333.7$ reported by operational IFS of ECMWF~\cite{tigger_data}. In~\cite{can_ai_beat_nwp}, researchers conjectured that `a number of fundamental breakthroughs are needed' before AI-based methods can beat NWP.

\begin{figure*}
\centering
\includegraphics[width=9cm]{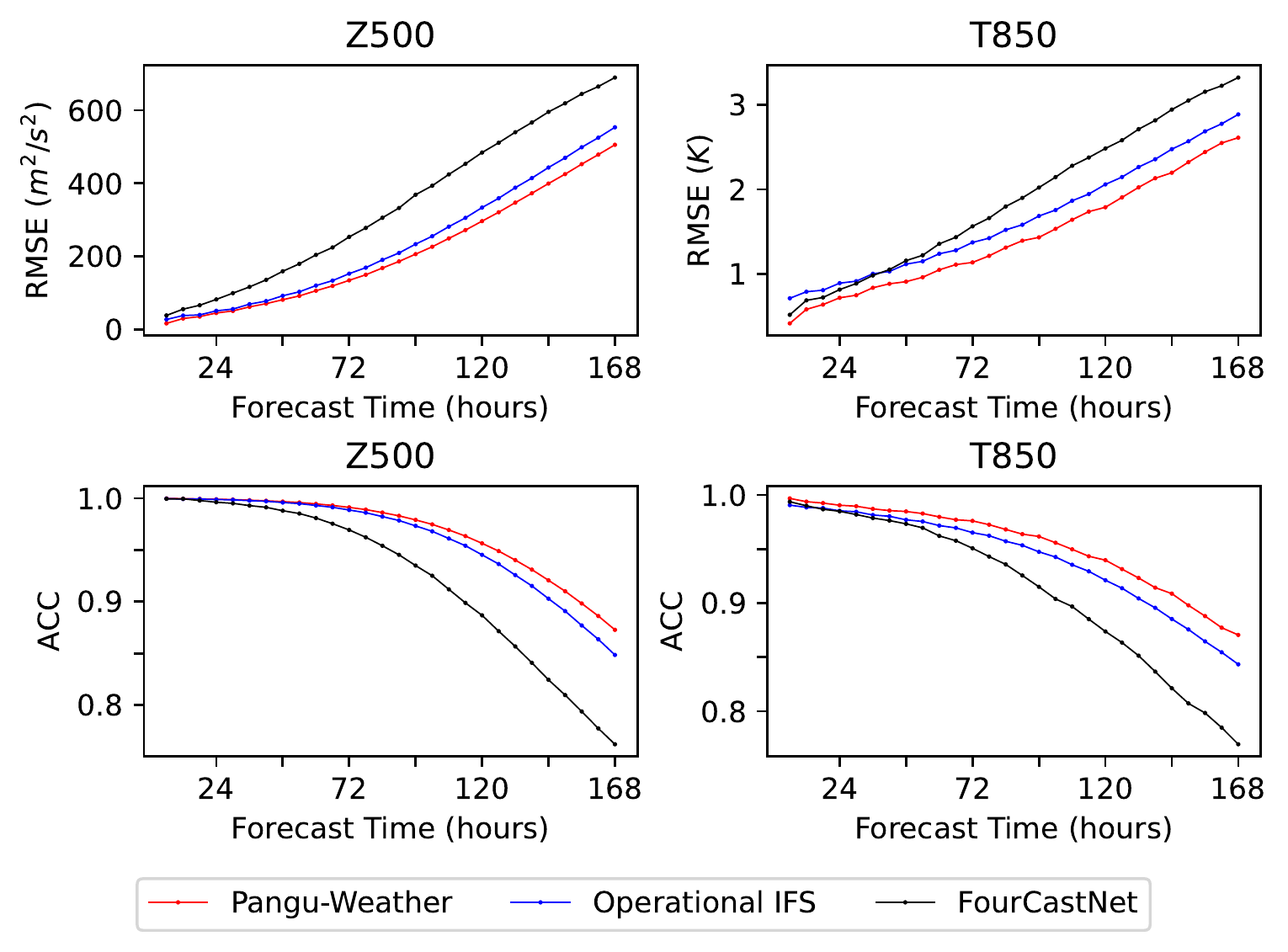}\hfill
\includegraphics[width=9cm]{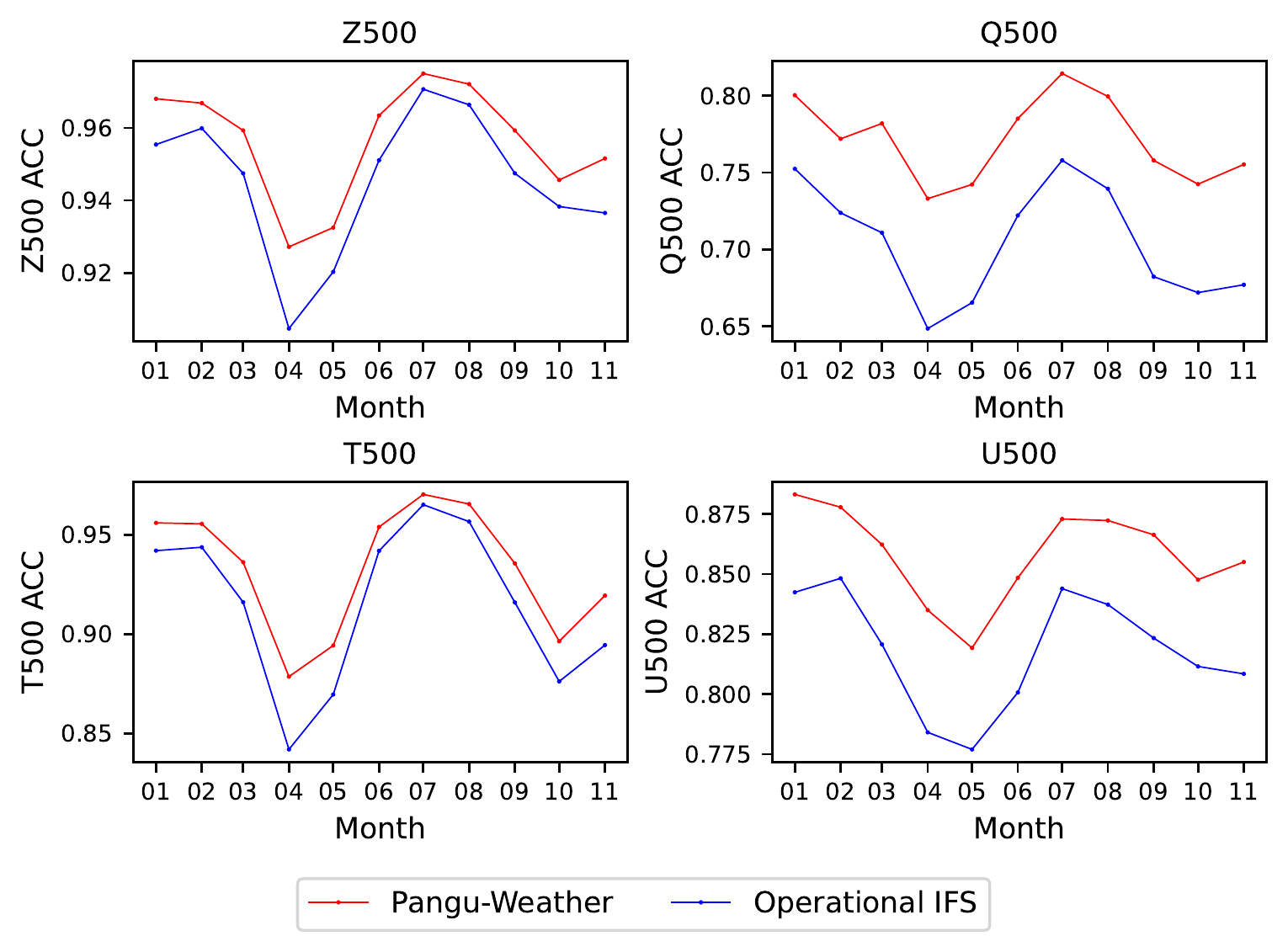}\\
\includegraphics[width=9cm]{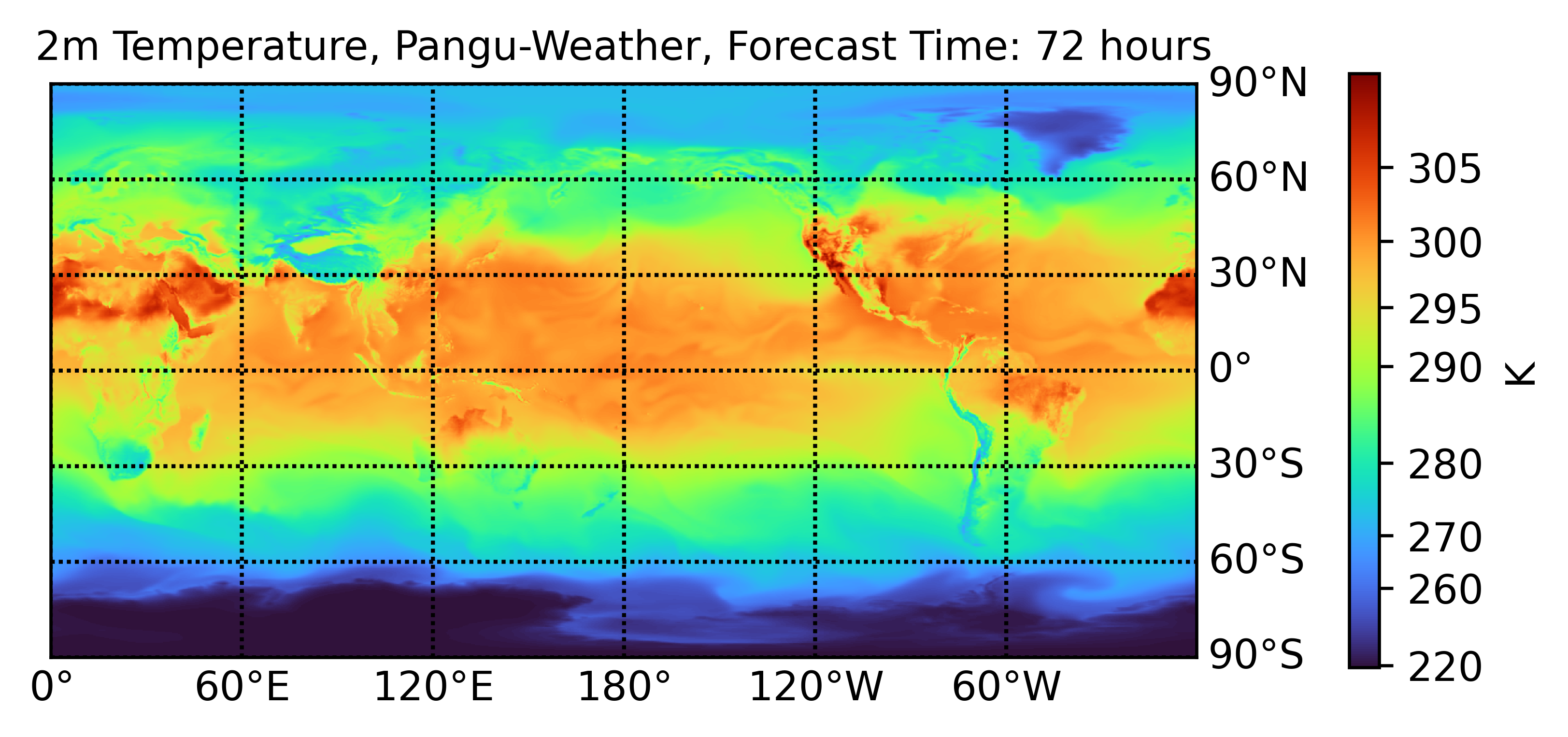}\hfill
\includegraphics[width=9cm]{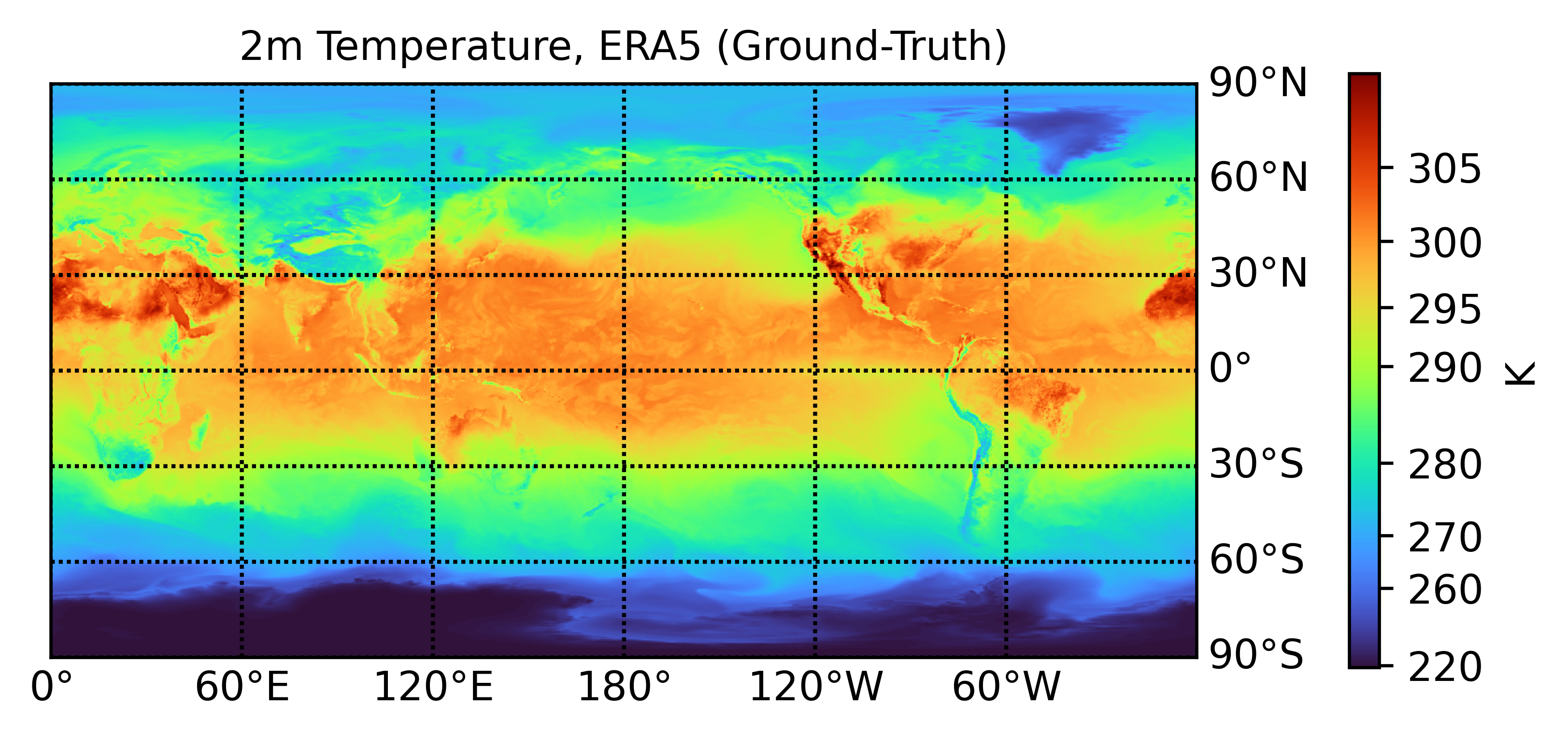}\\
\includegraphics[width=9cm]{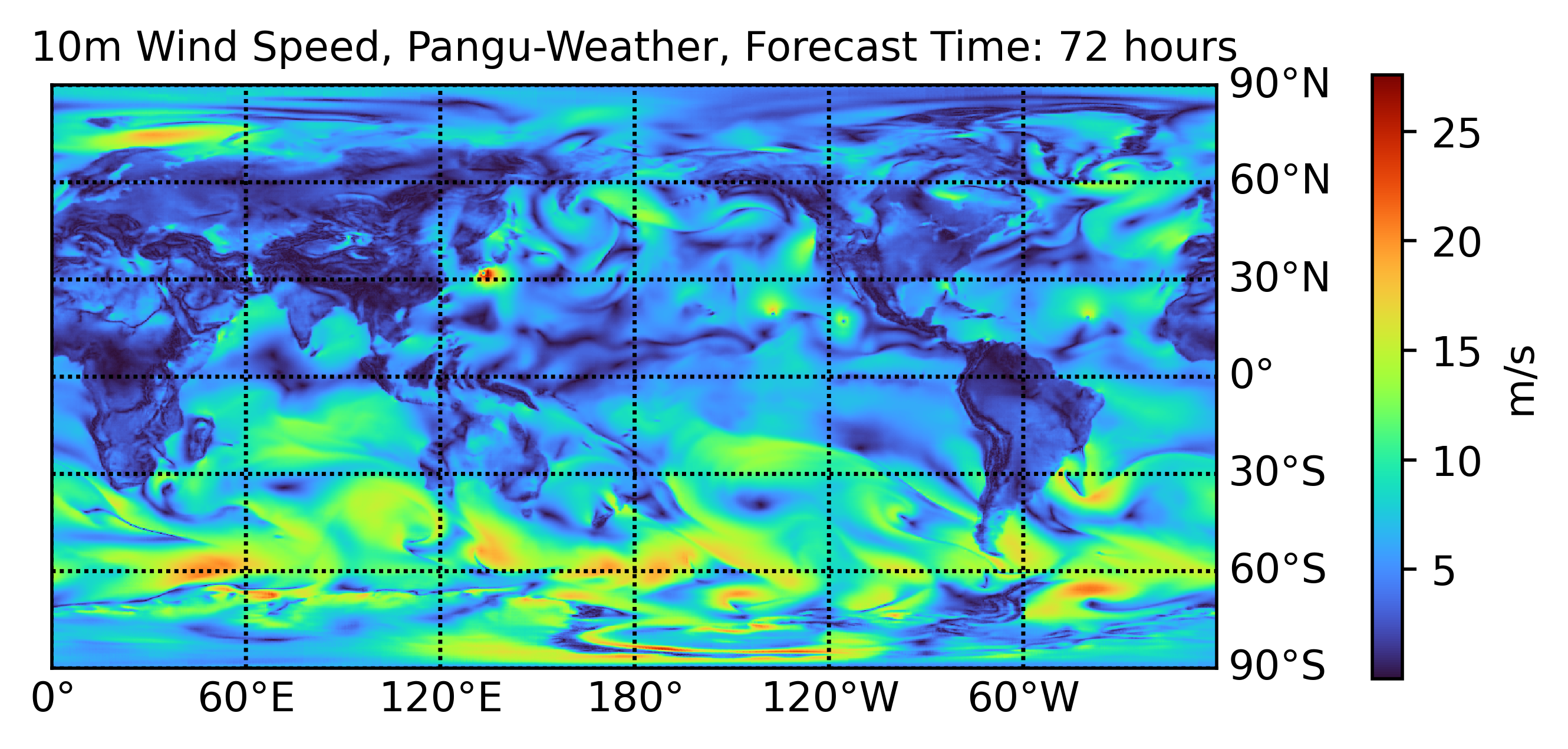}\hfill
\includegraphics[width=9cm]{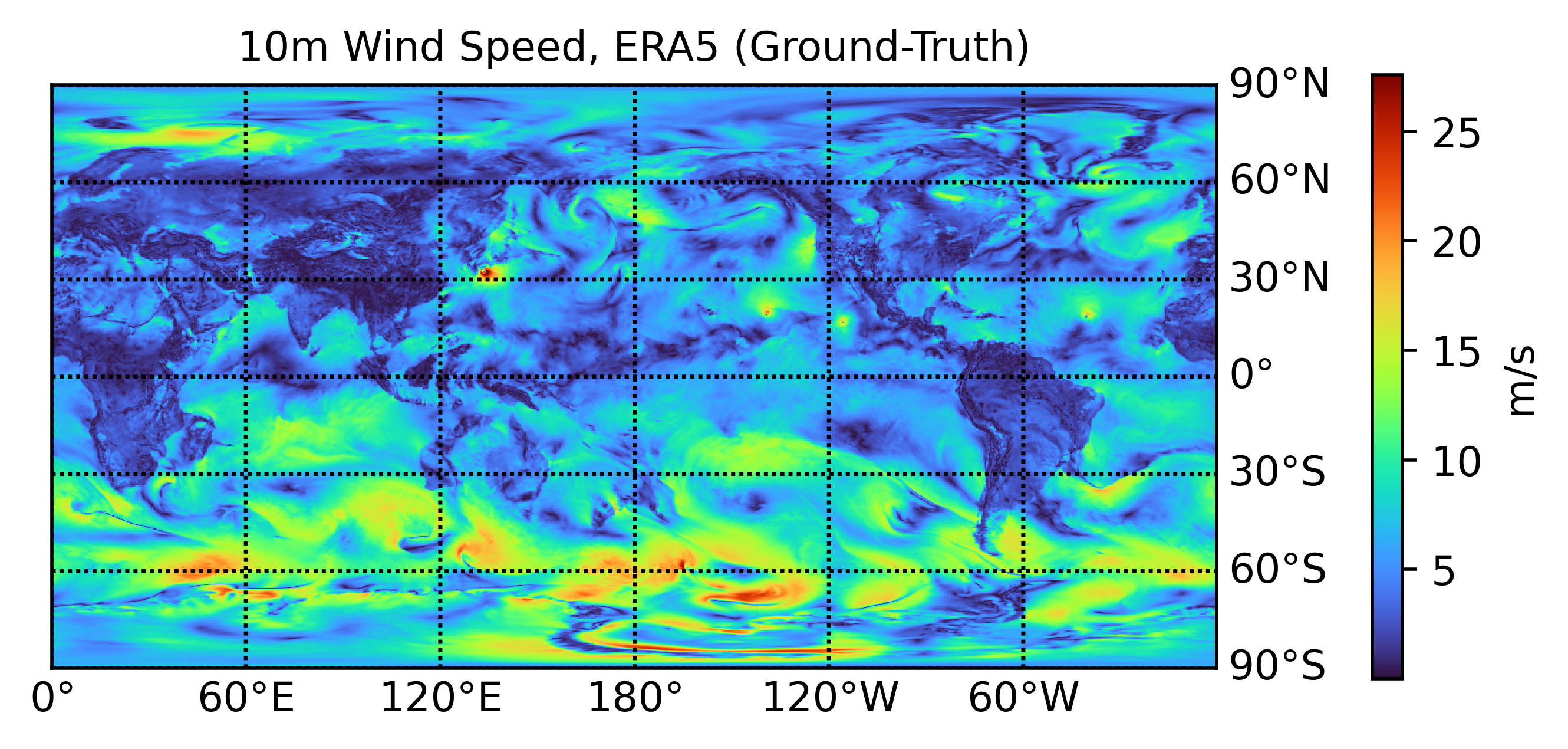}\\
\includegraphics[width=9cm]{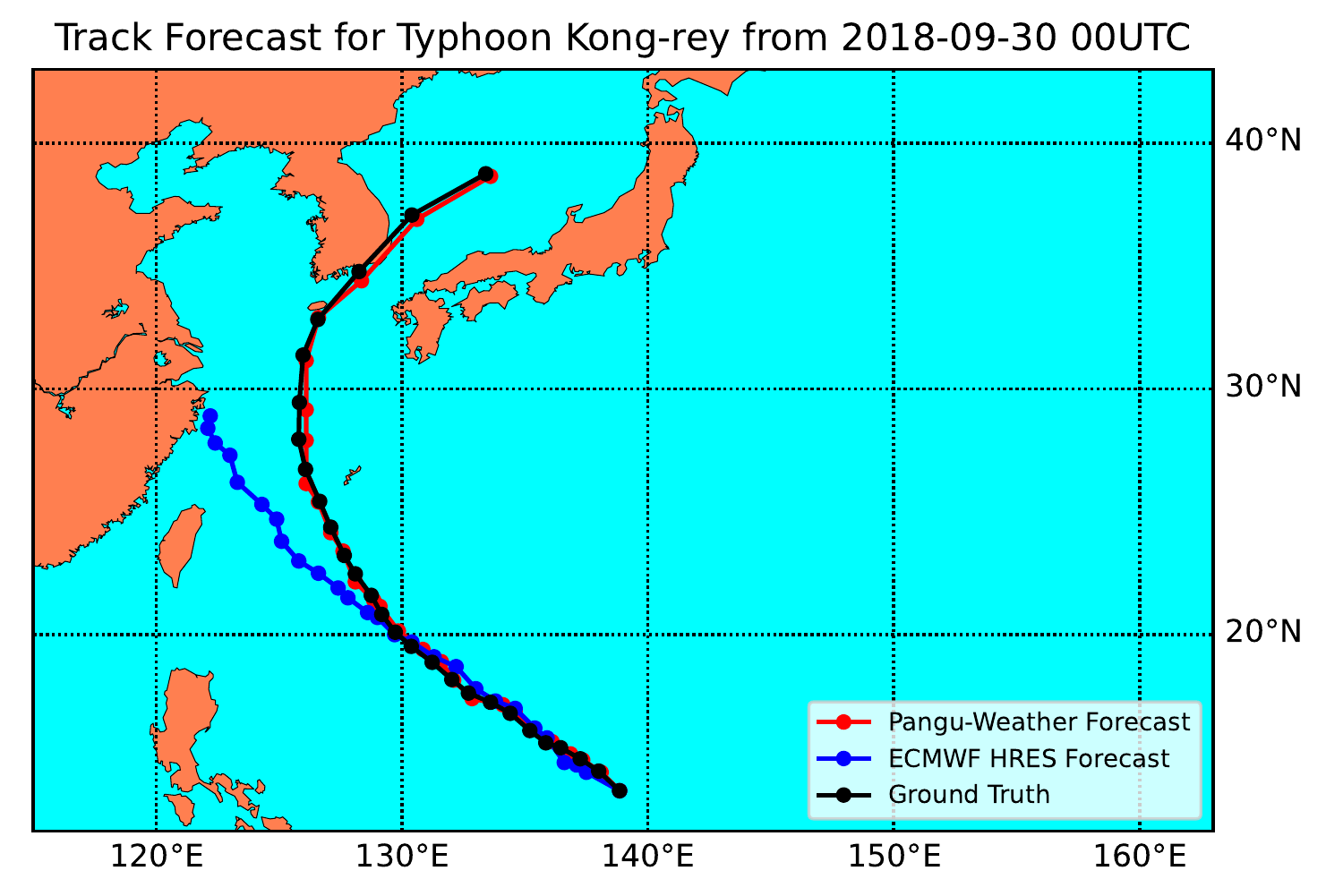}\hfill
\includegraphics[width=9cm]{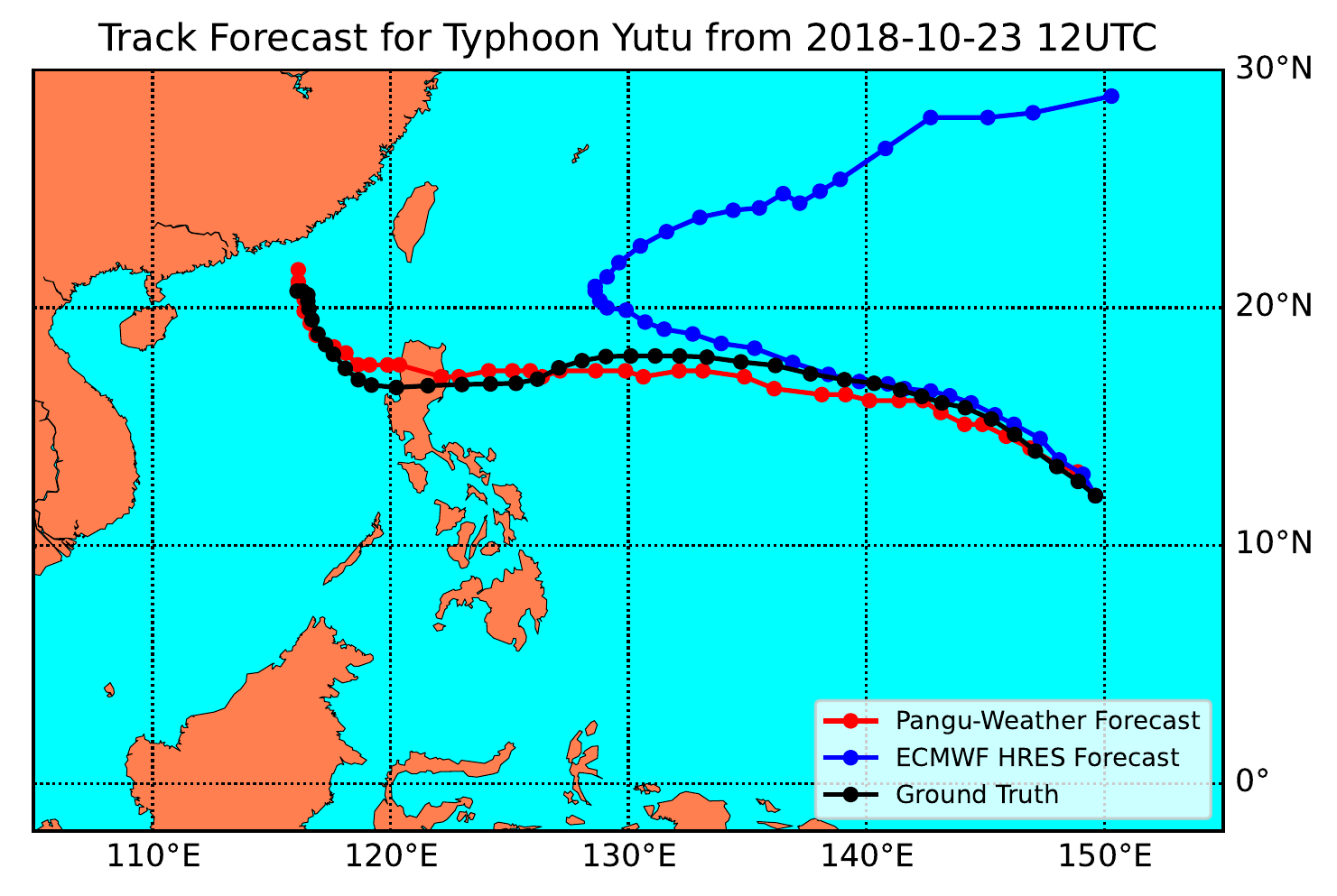}
\caption{A showcase of Pangu-Weather's forecast results. \textbf{Top}: Pangu-Weather claims significant advantages over operational IFS (NWP) and FourCastNet (AI-based) in terms of forecast accuracy (i) of different factors ($500\mathrm{hPa}$ geopotential, Z500, and $850\mathrm{hPa}$ temperature, T850) and (ii) with respect to different months in year. \textbf{Middle}: visualization of Pangu-Weather's $3$-day forecast of 2m temperature (T2M) and 10m wind speed at 00:00 UTC, September 1st, 2018, with comparison to the ERA5 ground-truth. \textbf{Bottom}: Pangu-Weather produces more accurate tracking for two tropical cyclones in 2018, \textit{i.e.}, Typhoon Kong-rey (2018-25) and Yutu (2018-26). Specifically, Pangu-Weather predicts the correct path of Yutu (\textit{i.e.}, it goes to the Philippines) $48$ hours earlier than the ECMWF-HRES forecast.}
\label{fig:overview}
\end{figure*}

The breakthrough comes much earlier than they thought. In this paper, we present Pangu-Weather, a powerful AI-based weather forecast system that, \textbf{for the first time}, surpasses existing NWP methods (and, of course, AI-based methods) in terms of prediction accuracy of all factors. The test is performed on the 5th generation of ECMWF reanalysis (ERA5) data. We download $43$ years (1979--2021) of global weather data, among which we use the 1979--2017 data for training, the 2019 data for validation, and the $2018$, $2020$, $2021$ data for testing. We choose $13$ pressure levels, each with $5$ important variables (\textit{i.e.}, geopotential, specific humidity, temperature, $u$-component and $v$-component of wind speed), and the surface level with $4$ variables (\textit{i.e.}, $2\mathrm{m}$ temperature, $u$-component and $v$-component of $10\mathrm{m}$ wind speed, and mean sea-level pressure).

Some key results are summarized in Figure~\ref{fig:overview}. \textbf{Quantitatively}, Pangu-Weather outperforms all existing weather forecast systems. In particular, with a single-member forecast, Pangu-Weather reports an RMSE of $5$-day Z500 forecast of $296.7$, significantly better than the operational IFS~\cite{tigger_data} and the previous best AI-based method (\textit{i.e.}, FourCastNet~\cite{fourcastnet}) which reported $333.7$ and $462.5$, respectively. In addition, the inference cost of Pangu-Weather is merely $1\rm{,}400\mathrm{ms}$ on a single GPU, more than $10000\times$ faster than operational IFS and on par with FourCastNet~\cite{fourcastnet}. \textbf{Qualitatively}, Pangu-Weather not only shows high-resolution ($0.25^\circ\times0.25^\circ$) visualization maps (\textit{e.g.}, for temperature and wind speed), but also offers high-quality extreme weather forecast (\textit{e.g.}, for tropical cyclone tracking).

The technical contribution of Pangu-Weather is two-fold. \textbf{First}, we integrate height information (offered by different pressure levels) into a new dimension, so that the input and output of deep neural networks are in 3D forms. We further design a 3D Earth-specific transformer (3DEST) architecture to process 3D data. Our experiments show that, although 3D data require heavier computational overhead (in particular, the large memory costs obstacle us from using full observation elements and very deep network architectures), 3D models can better capture the intrinsic relationship between different pressure levels and thus yield significant accuracy gain beyond the 2D counterparts. \textbf{Second}, we apply a hierarchical temporal aggregation algorithm that involves training a series of models with increasing forecast lead times (\textit{i.e.}, $1$-hour, $3$-hour, $6$-hour, and $24$-hour forecast). Hence, in the testing stage, the number of iterations needed for medium-range (\textit{e.g.}, $5$-day) forecast is largely reduced and, consequently, the cumulative forecast errors are alleviated. Compared to previous methods (\textit{e.g.}, FourCastNet~\cite{fourcastnet} applied plain temporal aggregation with recurrent optimization), our strategy is easier to implement, more stable during training, and achieves much higher medium-range forecast accuracy.

The Pangu-Weather system is built upon a GPU cluster of Huawei Cloud with $192$ NVIDIA Tesla-V100 GPUs. Each single forecast model is trained for $100$ epochs which take around $15$ days. To maximally support large neural networks, we use a batch size of $1$ on each GPU, \textit{i.e.}, the overall batch size is $192$. With diagnostic studies, we notice that the forecast accuracy continuously goes up with a larger amount of training data and/or a longer training procedure -- $100$ epochs, the maximum budget that we can use, are actually insufficient for the training procedure to arrive at full convergence. That said, the community can wait for more data (including increasing the time and/or spatial resolutions) or use more powerful computational device to improve AI-based weather forecast. The trend is similar to establishing large-scale pre-trained models in other AI scopes, \textit{e.g.}, computer vision~\cite{moco,beit}, natural language processing~\cite{bert,gpt3}, cross-modal understanding~\cite{clip}, and beyond.

Overall, the contribution of this paper are summarized in the following three aspects:
\begin{itemize}
\item We end the debate on whether AI-based methods can surpass NWP for global weather forecast. We establish a deep learning framework that, for the first time, surpasses operational IFS in terms of all weather factors and all forecast times from one hour to one week, meanwhile enjoying a very fast inference speed and a high spatial resolution of $0.25^\circ\times0.25^\circ$.
\item Technically, we reveal several key issues that significantly improve forecast accuracy, namely, (i) using a 3D deep network to integrate height information, and (ii) applying hierarchical temporal aggregation to alleviate cumulative forecast errors. Arguably, these techniques will be more effective in the future with more powerful computational device and higher-quality training data.
\item We show that Pangu-Weather can easily transfer the ability of deterministic forecast to downstream scenarios such as extreme weather forecast and large-member ensemble forecast, where timeliness is guaranteed by its fast inference speed.
\end{itemize}

The remainder of this paper is organized as follows. Section~\ref{preliminaries} formulates the problem and briefly reviews previous work, based on which we demonstrate our technical insights. Section~\ref{approach} elaborates the Pangu-Weather system with algorithmic designs and implementation details. Section~\ref{results} shows generic forecast results and investigates two specific scenarios, namely, extreme weather forecast and large-member ensemble forecast. Section~\ref{conclusions} draws conclusions and reveals future directions.

\section{Preliminaries and Insights}
\label{preliminaries}

\subsection{Problem Setting and Notations}
\label{preliminaries:setting}

Most weather forecast systems were built upon the analysis or reanalysis beyond observation data. The reanalysis datasets are considered the best known estimation~\cite{era5_bias_1,era5_bias_2} for most atmospheric variables except for some factors like precipitation. Throughout this paper, we make use of the ERA5 dataset, \textit{i.e.}, the 5th generation of ECMWF reanalysis data~\cite{era5_data}. The ERA5 data have four dimensions, namely, latitude and longitude, pressure levels (for height) and time. We can choose an arbitrary number of weather factors (\textit{e.g.}, geopotential, temperature, \textit{etc.}), but do not count them toward a new dimension. With a total size of over $2\mathrm{PB}$, the dataset is split into 2D (latitude and longitude) slices to ease downloading. That said, given a time point (hourly within the past $60$ years), a pressure level (or Earth's surface), and a weather factor, one can download a matrix representing the specified global reanalysis data. We denote the overall ERA5 data as $\mathbf{A}$, and we use superscripts to refer to specific weather factors and pressure levels, and subscripts to indicate spatiotemporal coordinates, \textit{e.g.}, $\mathbf{A}_t^\mathrm{T850}$ stands for the global temperature data (a matrix) at time $t$ and a height of $850\mathrm{hPa}$ and $\mathbf{A}_{x,y,t}^\mathrm{Z500}$ the geopotential data at position $(x,y)$, time $t$, and a height of $500\mathrm{hPa}$ -- note that $\mathbf{A}_{x,y,t}^\mathrm{Z500}$ is a single number.

Based on the ERA5 data, the problem of weather forecast is clearly defined: given an initial time $t_0$, the algorithm shall make use of all historical weather data (\textit{i.e.}, $\mathbf{A}_t^\ast$ for $t\leqslant t_0$) to predict future weather data (\textit{i.e.}, $\mathbf{A}_t^\ast$ for $t>t_0$), where $\ast$ stands for all factors. Before starting a survey on existing methods, we first note that the resolution of weather data is large due to the following facts. First, there are $37\times21+262$ observation factors in total ($37$ pressure levels, each of which has $21$ weather variables, and a surface that has $262$ variables), and it is believed that different elements can impact each other (\textit{e.g.}, temperature is highly correlated to geopotential). Second, ERA5 provides about $60$ years of hourly observation data, \textit{i.e.}, the scale of time axis is over $10^5$. Third, the spatial resolution is $0.25^\circ\times0.25^\circ$, implying that each frame of global weather data is of $1440\times720$ numbers (\textit{i.e.}, `pixels' or `voxels' if the data is to be processed by deep neural networks). As we shall see later, the high complexity has raised serious concerns on computational costs for both NWP and AI-based methods.

Based on the above definition, a weather forecast system is described as a mathematical function $f(\cdot)$ applied on $\mathbf{A}_t^\ast$. There are mainly two lines of research for weather forecast, which we follow the convention~\cite{can_ai_beat_nwp} to refer to them as NWP and AI-based methods.

\begin{table}[]
\centering
\setlength{\tabcolsep}{0.08cm}
\begin{tabular}{|c|l|}
\hline
Terms & Definition in this paper \\
\hline\hline
system & The entire algorithm for end-to-end weather forecast \\
model & A deep network that produces one-time prediction \\
\hline\hline
initial time & The time point that weather forecast is made at \\
forecast time & The time gap between observation and desired forecast \\
lead time & The time gap between input and output of one model \\
\hline
spacing & The minimum forecast time in a forecast system \\
range & The maximum forecast time in a forecast system \\
variable & An observed weather factor, \textit{e.g.}, 2m temperature \\
parameter & A learnable value in deep networks \\
\hline\hline
$x,y$ & Horizontal coordinate (latitude \& longitude) \\
$t$ & Temporal coordinate (time point) \\
$\Delta t$ & Lead time added to $t$ \\
$h$ & Height (in pressure level, \textit{e.g.}, $500\mathrm{hPa}$) \\
\hline
$\mathbf{A}$ & The overall weather data (\textit{e.g.}, ERA5) \\
$\mathbf{A}_{x,y,t}^{\mathrm{T}h}$ & Temperature at position $(x,y)$, time $t$, height $h$ \\
$\mathbf{A}_{t}^\ast$ & All variables (all positions and heights) at time point $t$ \\
$\hat{\mathbf{A}}_{t+\Delta t}^\ast$ & The forecast results at time point $t+\Delta t$ \\
\hline
\end{tabular}
\caption{A summary of terminologies and notations used in this paper. In this work, we name the proposed system as Pangu-Weather and the proposed model architecture as 3D Earth-specific transformers (3DEST).}
\label{tab:definition}
\end{table}

\subsection{NWP Methods}
\label{preliminaries:nwp}

The first line is the conventional numerical weather prediction (NWP) methods that approximate $f(\cdot)$ using simulation. Starting with initial weather states, a set of partial differential equations (PDEs) are established to simulate different physical processes such as thermodynamics equations, N-S equations, continuous equations, \textit{etc}~\cite{nwp_evolution, origins_nwp, weather_forecast_theory}. To solve the PDEs, the atmospheric states are partitioned into discrete grids. Intuitively, reducing the spacing of grids leads to a larger number of grids and a higher spatial resolution of weather forecast, and also increases the computational costs of simulation. Currently, the spatial resolution is highly limited by the power of supercomputers. To accelerate computation, more approximation approaches were introduced, including (i) interpolation, which first performs low-resolution simulation and then estimates in-grid weather states, and (ii) parameterization~\cite{parameterization_scheme}, which uses an approximate function to solve very complex weather processes -- typical examples include the parameterization for cloud~\cite{parameterization_cloud_example,parameterization_cloud_example_2,parameterization_cloud_example_3} and convection~\cite{parameterization_convection_example,parameterization_convection_example_2}.

Prior to this work, NWP methods contribute overall the highest prediction accuracy, but they are still troubled by the super-linearly increasing computational overhead~\cite{nwp_evolution,ecmwf_speed}, especially when the amount of observation data keeps growing and it is difficult to perform efficient parallelization for NWP methods~\cite{nwp_gpu_challenge}. The slowness of NWP not only weakens the timeliness of operational IFS systems (\textit{e.g.}, most such systems can only update prediction several times a day), but also restricts the number of ensemble members (\textit{i.e.}, a set of individual prediction results for ensemble), hence weakening the diversity and accuracy of probabilistic weather forecast. In addition, the formulae used by NWP methods inevitably introduce approximation and computational errors~\cite{nwp_uncertainty_book,model_uncertainty_review} which can augment with either iteration or incomplete or inaccurate analysis data~\cite{data_assimilation_review}. It thus brings major challenges to maintain NWP methods with a complex PDE system that takes more and more factors into consideration.

\subsection{AI-based Methods}
\label{preliminaries:dl}

To alleviate the above burden, researchers started the second line that investigates AI-based methods for weather forecast. The cutting edge technology of AI lies in deep learning~\cite{lecun2015deep}, a branch of machine learning, assuming that the complex function (\textit{i.e.}, $f(\cdot)$) can be directly learned from abundant training data without knowing the actual physical procedure and/or formulae. Most often, $f(\cdot)$ appears as a deep neural network which is often written as $f(\cdot;\boldsymbol{\theta})$ where $\cdot$ is a placeholder for input data and $\boldsymbol{\theta}$ denotes the learnable parameters. The network often contains a number of layers. Each of these layers has a large amount of learnable parameters, and these parameters are initialized as white noise and optimized by back-propagating prediction errors of the deep network. The most similar field to weather forecast is computer vision (CV) where image data appears in 2D/3D cubes. In the past decade, the CV community developed many effective network architectures (\textit{e.g.}, \cite{krizhevsky2012imagenet,he2016deep}, \textit{etc.}), and recently, they transplanted a kind of powerful architectures named transformers from natural language processing~\cite{vaswani2017attention} and developed the variants~\cite{vision_transformer,swin_transformer} that are capable of dealing with image data.

In the scope of weather forecast, AI-based methods were first applied in the scenarios where it is difficult to predict future weather data using the NWP methods, \textit{e.g.}, precipitation forecasting based on radar data~\cite{nowcasting_shi2015,now_casting_shi2017,nowcasting_agrawal2019,nowcasting_deepmind2021} or satellite data~\cite{nowcasting_sat_2019,nowcasting_sat_metnet}. The powerful expressive ability of deep neural networks led to the success in these data-driven environments, which further encouraged researchers to delve into the scenarios that the NWP methods are troubled by enormous computational overhead, \textit{e.g.}, direct medium-range weather forecast~\cite{weatherbench,graph_weather,fourcastnet,swinvrnn,weyn_new} that consumed most of the computational resources of weather forecast centers in the past decade.


This paper investigates medium-range weather forecast. NWP and AI-based methods have been competing in this scenario, where NWP methods led in forecast accuracy~\cite{can_ai_beat_nwp} and resolution, while AI-based methods showed their advantages in efficiency (\textit{e.g.}, the inference speed is orders of magnitude faster than the NWP methods~\cite{fourcastnet,can_ai_beat_nwp}). Prior to 2022, AI-based methods cannot achieve the horizontal resolution of $0.25^{\circ}\times0.25^{\circ}$ as NWP methods can. Recently, FourCastNet~\cite{fourcastnet} improved the resolution to $0.25^{\circ}\times0.25^{\circ}$, but the forecast accuracy (\textit{e.g.}, in terms of RMSE or ACC), is still inferior to operational IFS even after a large-member ensemble was performed. The disadvantages in forecast accuracy and interpretability, especially in extreme weather events, hinder the applications of AI-based methods. Consequently, AI-based methods mostly play the role of fast surrogate models for medium-range weather forecast.

\subsection{Insights}
\label{preliminaries:insights}

We briefly analyze the reasons why AI-based (specifically, deep learning based) methods fell behind NWP methods in terms of prediction accuracy. There are mainly two aspects, summarized as follows.

\textbf{First}, weather forecast shall take high-dimensional (\textit{e.g.}, 3D spatial with 1D time), anisotropic data into consideration, yet existing AI-based methods~\cite{weatherbench,graph_weather,weyn_new,swinvrnn,fourcastnet} often worked on 2D (latitude and longitude) data. This brings two-fold disadvantages. On the one hand, the spacing and distribution of atmospheric states and the relationship between atmospheric patches change rapidly across pressure levels, making it difficult for 2D models to adapt to different situations. On the other hand, many weather processes (\textit{e.g.}, radiation, convection, \textit{etc.}) can only be completely formulated in the 3D space, and thus 2D models cannot make use of such important patterns.

\textbf{Second}, medium-range weather forecast can suffer from cumulative forecast errors when the model is called too many times. As an example, FourCastNet~\cite{fourcastnet} trained a base model for $6$-hour forecast, so that performing a $7$-day forecast required executing the model $28$ times iteratively. Compared to the case in NWP methods, such errors can grow rapidly because AI-based methods often do not consider real-world constraints (\textit{e.g.}, formulated by the PDEs). According to the results in FourCastNet, the forecast error often grows super-linearly with time. Note that FourCastNet applied a specialized method for reducing iteration error, but the actual gain is somewhat limited.

Summarizing the above factors, we come up with the insights that one shall try to \textbf{increase the dimensionality of data} and \textbf{reduce the number of iterations} for more accurate medium-range weather forecast. However, this encounters difficulties in computational overhead because the weather data is very large (see Section~\ref{preliminaries:setting}). In the next part, we will elaborate a method built upon a tradeoff between accuracy and efficiency -- in brief, we use 3D (latitude, longitude, and height) data as input and output, and train a few individual models for different prediction time gaps to maximally reduce the maximum number of iterations called for medium-range forecast.

\section{Methodology}
\label{approach}

\subsection{Overview}
\label{approach:overview}

\begin{figure*}[!t]
\centering
\includegraphics[width=18cm]{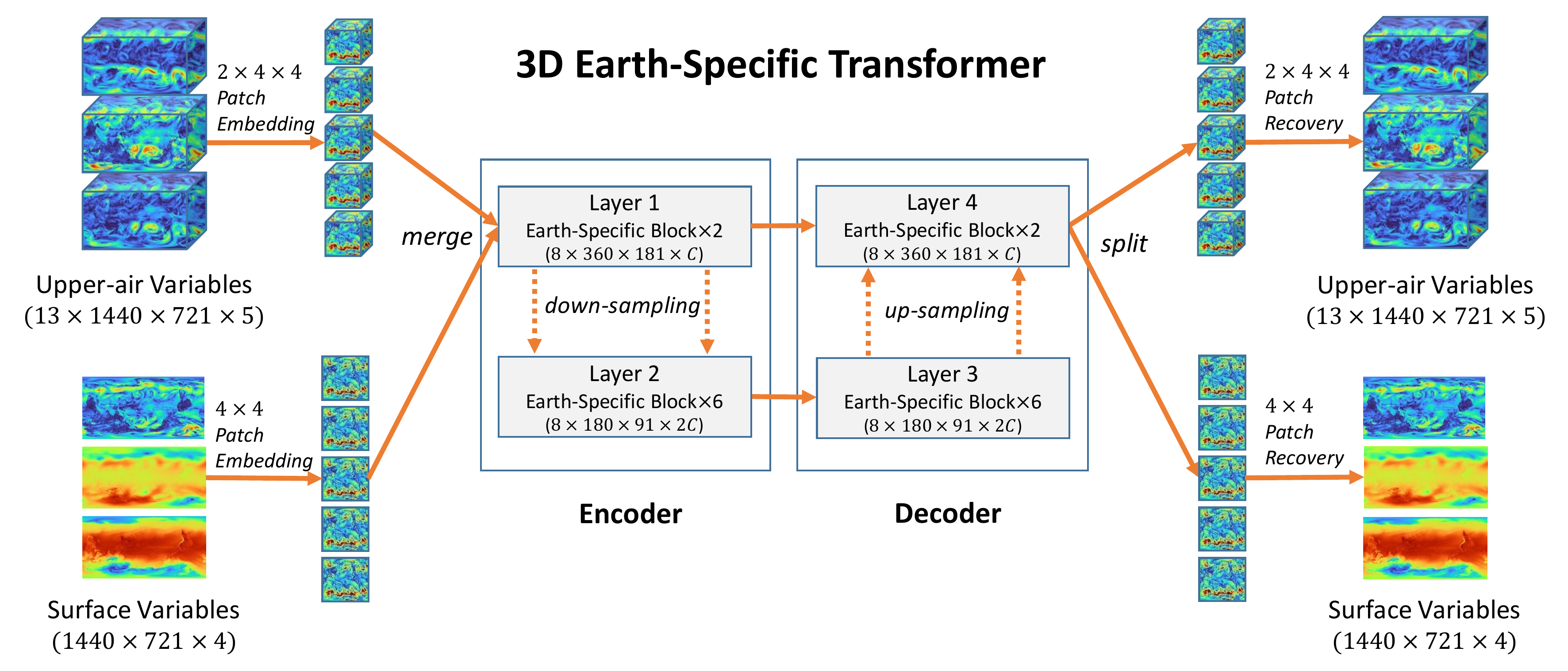}
\caption{An overview of the 3D Earth-specific transformer (3DEST). Based on the standard encode-decoder design, we (i) adjust the shifted-window mechanism and (ii) apply an Earth-specific positional bias -- see the main texts for details.}
\label{fig:3d_transformer}
\end{figure*}

Based on the above insights, we present our system, termed Pangu-Weather, for fast and accurate global weather forecast. As an AI-based method, it surpasses the accuracy of conventional NWP methods for the first time, meanwhile enjoying a very fast inference speed.

The core part of Pangu-Weather is a set of deep neural networks trained on $39$ years of global weather data -- we elaborate data preparation and the pre-training task in Section~\ref{approach:data}. The key to reduce the accuracy loss is two-fold, namely (i) using a 3D Earth-specific transformer (3DEST) to model the 3D atmosphere effectively -- see Section~\ref{approach:architecture}, and (ii) applying a hierarchical temporal aggregation strategy (\textit{i.e.}, training a few models with various lead times) to alleviate cumulative forecast errors -- see Section~\ref{approach:temporal}. The Pangu-Weather system can be applied to generic or specific forecast scenarios, as we shall see in Section~\ref{results}.

\subsection{Data Preparation and the Pre-training Task}
\label{approach:data}

We download the ERA5 dataset~\cite{era5_data, era5_pressure, era5_surface} from the official website\footnote{\textsf{https://cds.climate.copernicus.eu/} offered by Copernicus Climate Data (CDS).} for training and evaluating Pangu-Weather. It contains global, hourly reanalysis data for the past $60$ years. The observation data and the prediction of numerical models are blended into reanalysis data using numerical assimilation methods, providing a high-quality benchmark for global weather forecast. Following the existing methods~\cite{weatherbench,fourcastnet,graph_weather}, we train our models on a subset of ERA5 -- in particular, we use the 1979--2017 ($39$ years of) data for training, the 2019 data for validation, and the 2018, 2020, 2021 data for testing.

We make use of observation data of every single hour so that the algorithm can perform hourly prediction. We keep the highest spatial resolution available in ERA5, namely, $0.25^{\circ}\times0.25^{\circ}$ on Earth's sphere, resulting in an input resolution of $1440\times721$ ($1440$ for longitude and $721$ for latitude -- note that the northmost and southmost data do not overlap). The largest difference between our method and the prior works lies in that we formulate height information (represented as pressure levels) into the 3rd spatial dimension. To reduce computational costs, we follow~\cite{weatherbench} to choose $13$ pressure levels (\textit{i.e.}, $50\mathrm{hPa}$, $100\mathrm{hPa}$, $150\mathrm{hPa}$, $200\mathrm{hPa}$, $250\mathrm{hPa}$, $300\mathrm{hPa}$, $400\mathrm{hPa}$, $500\mathrm{hPa}$, $600\mathrm{hPa}$, $700\mathrm{hPa}$, $850\mathrm{hPa}$, $925\mathrm{hPa}$, and $1000\mathrm{hPa}$), from a total of $37$ levels, plus Earth's surface. To fairly compare with the online version of ECMWF control forecast, we choose to predict the factors published in the TIGGE dataset~\cite{tigger_data}, namely, five upper-air atmospheric variables (\textit{i.e.}, geopotential, specific humidity, temperature, $u$-component and $v$-component of wind speed) and four surface weather variables (\textit{i.e.}, $2\mathrm{m}$ temperature, $u$-component and $v$-component $10\mathrm{m}$ wind speed, and mean sea level pressure). In addition, three constant masks (\textit{i.e.}, the topography mask, land-sea mask and soil type mask) are added to the input of surface variables.

The pre-training task is straightforward, \textit{i.e.}, asking the model to predict the future weather given historical observation data. Technically, this involves sampling a time point $t$ (\textit{i.e.}, date and hour) from the dataset and specifying a prediction gap $\Delta t$, so that the model, $f(\cdot;\boldsymbol{\theta})$, takes $\mathbf{A}_t^\ast$ as input and predicts $\hat{\mathbf{A}}_{t+\Delta t}^\ast$, with the goal of approaching $\mathbf{A}_{t+\Delta t}^\ast$. In the context of deep learning, $f(\cdot;\boldsymbol{\theta})$ appears as a differentiable function so that the difference between $\mathbf{A}_{t+\Delta t}^\ast$ and $\hat{\mathbf{A}}_{t+\Delta t}^\ast$ is computed and back-propagated to update the parameters, $\boldsymbol{\theta}$. The technical details, including the design of $f(\cdot;\boldsymbol{\theta})$ and the choice of $\Delta t$ values, are to be elaborated in the following subsections.

\subsection{3D Earth-Specific Transformer}
\label{approach:architecture}

This part describes the design of $f(\cdot,\boldsymbol{\theta})$. We name it as a 3D Earth-specific transformer (3DEST). The overall architecture of 3DEST is illustrated in Figure~\ref{fig:3d_transformer}. It is a variant of vision transformer~\cite{vision_transformer} with input and output being 3D weather states at a specified time point. For a single model, the lead time between input and output states is fixed, \textit{e.g.}, $\Delta t$ equals to $6$ hours. We achieve any-time weather forecast by aggregating multiple models with different lead times, as elaborated in the next subsection.

There are two sources of input and output data, namely, upper-air variables and surface variables. The former involves $13$ pressure levels, and they combined offer a $13\times1440\times721\times5$ data cube. The latter contains a $1440\times721\times4$ cube. These parameters are first embedded from the original space into a $C$-dimensional latent space. A common technique in computer vision named patch embedding is used for dimensionality reduction. For the upper-air part, the patch size is $2\times4\times4$, so that the embedded data has a shape of $7\times360\times181\times C$. For the surface variables, the patch size is $4\times4$, so that the embedded data has a shape of $360\times181\times C$. These two data cubes are then concatenated along the first (height) dimension to yield a $8\times360\times181\times C$ cube. The cube is then propagated through a standard encoder-decoder architecture with $8$ encoder layers and $8$ decoder layers. The output of decoder is still a $8\times360\times181\times C$ cube, which is projected to the original space with patch recovery, producing the desired output. Below, we describe the technical details of each component.

\textbf{Patch embedding and patch recovery.}\quad
We follow the standard vision transformer to use a linear layer with GeLU activation for this purpose. In our implementation, a patch has $2\times4\times4$ pixels for upper-air variables and $4\times4$ for surface variables. The stride of sliding windows is the same as patch size, and necessary zero-value padding is added when the data size is indivisible by the patch size. The number of parameters for patch embedding is $(4\times4\times2\times5)\times C$ for upper-air variables and $(4\times4\times4)\times C$ for surface variables. Patch recovery performs the opposite operation, but it does not share parameters with patch embedding.

\textbf{The encoder-decoder architecture.}\quad
The data size remains unchanged ($8\times360\times181\times C$) for the first $2$ encoder layers, while for the next $6$ layers, the horizontal dimensions are reduced by a factor of $2$ and the number of channels is doubled, resulting in a data size of $8\times180\times91\times2C$. The decoder part is symmetric to the encoder part, with the first $6$ decoder layers sized $8\times180\times91\times2C$ and the next $2$ layers sized $8\times360\times181\times C$. The outputs of the 2nd encoder layer and the 7th decoder layer are concatenated along the channel dimension. Down-sampling and up-sampling operations connect the adjacent layers of different resolutions, and we follow the implementation of Swin transformers~\cite{swin_transformer}. For down-sampling, we merge four tokens into one (the feature dimensionality increases from $C$ to $4C$) and perform a linear layer to reduce the dimensionality to $2C$. For up-sampling, the reverse operations are performed.

\begin{figure*}[!t]
\centering
\includegraphics[width=18cm, trim={3cm 0 0 0},clip]{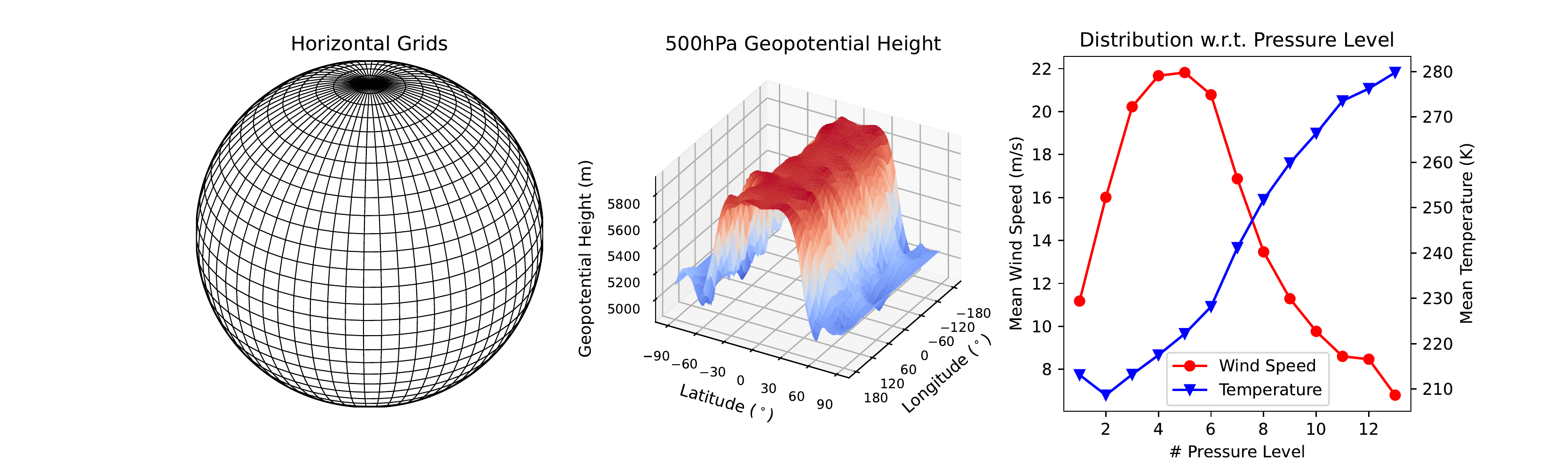}
\caption{The motivation of using an Earth-specific positional bias. \textbf{Left}: the horizontal map corresponds to an uneven spatial distribution on Earth's sphere. \textbf{Middle}: the geopotential height is closely related to the latitude. \textbf{Right}: the mean wind speed and temperature are closely related to the height (formulated as pressure levels).}
\label{fig:irregular_grid}
\end{figure*}

\textbf{3D Earth-specific transformer blocks.}\quad
Each encoder and decoder layer is a 3D Earth-specific transformer (3DEST) block. It is similar to the standard vision transformer block~\cite{vision_transformer} but specifically designed to align with Earth's geometry. To further reduce computational costs, we inherit the window-attention mechanism~\cite{swin_transformer} to partition the feature maps (either $8\times360\times181$ or $8\times180\times91$ -- the last dimension is omitted) into windows, and each window contains up to $2\times12\times6$ tokens. The standard self-attention mechanism is applied within each window. The shifted-window attention mechanism is applied, so that for every layer, the grid partition differs from the previous one by half window size\footnote{Note that, along the longitude dimension, the leftmost and rightmost indices are actually close to each other. In the shifted-window mechanism, if half windows appear at both leftmost and rightmost positions, they are directly merged into one window.}. We refer the reader to the original paper~\cite{swin_transformer} for more details. The standard self-attention formula is written below:
\begin{equation}
\label{eqn:transformer}
\mathrm{Attention}(\mathbf{Q},\mathbf{K},\mathbf{V})=\mathrm{SoftMax}(\mathbf{Q}\mathbf{K}^\top/\sqrt{D}+\mathbf{B})\mathbf{V},
\end{equation}
where $\mathbf{Q}$, $\mathbf{K}$, and $\mathbf{V}$ are the query, key, and value vectors produced by the transformer block, respectively, $D$ is the feature dimensionality of $\mathbf{Q}$ and $\mathbf{K}$ (\textit{i.e.}, $C$ or $2C$), and $\mathbf{B}$ is the positional bias term.

\textbf{Earth-specific positional bias.}\quad
Swin transformer used a relative positional bias to represent the translation invariant component of attentions, where the bias is computed upon the relative coordinate of each window. For global weather forecast, however, the situation is a bit different. Each token corresponds to an absolute position on Earth's coordinate system and, since the map is a projection of Earth's sphere, the spacing between neighboring tokens can be different -- see Figure~\ref{fig:irregular_grid}. More importantly, some weather states are closely related to the absolute position. Examples of geopotential, wind speed, and temperature are shown in Figure~\ref{fig:irregular_grid}. To capture these properties, we modify $\mathbf{B}$ into an Earth-specific positional bias, termed $\mathbf{B}_\mathrm{ESP}$, adding a positional bias to each token based on its absolute (rather than relative) coordinate.

Mathematically, let the entire feature map have a spatial resolution of $N_\mathrm{pl}\times N_\mathrm{lat}\times N_\mathrm{lon}$ where $N_\mathrm{pl}$, $N_\mathrm{lat}$, and $N_\mathrm{lon}$ indicate the size along the axes of height (by pressure levels), latitude, and longitude, respectively. Swin transformer partitions these neurons into $M_\mathrm{pl}\times M_\mathrm{lat}\times M_\mathrm{lon}$ windows, and each window has a size of $W_\mathrm{pl}\times W_\mathrm{lat}\times W_\mathrm{lon}$. The Earth-specific position bias matrix contains $M_\mathrm{pl}\times M_\mathrm{lat}$ sub-matrices ($M_\mathrm{lon}$ does not appear because different longitudes share the same bias -- the longitude indices are cyclic and spacing is evenly distributed along this axis), each of which contains $W_\mathrm{pl}^2\times W_\mathrm{lat}^2\times(2W_\mathrm{lon}-1)$ learnable parameters. When the attention is computed between two units within the same window (Swin does not compute inter-window attentions), we use the window coordinate $(m_\mathrm{pl},m_\mathrm{lat},m_\mathrm{lon})$ to locate the corresponding bias sub-matrix ($m_\mathrm{lon}$ is not used), and then use the intra-window coordinates, $(h'_1,\phi'_1,\lambda'_1)$ and $(h'_2,\phi'_2,\lambda'_2)$, to call for the bias value at $(h'_1+h'_2\times W_\mathrm{pl},\phi'_1+\phi'_2\times W_\mathrm{lat},\lambda'_1-\lambda'_2+W_\mathrm{lon}-1)$ of the sub-matrix.


Applying the Earth-specific positional bias brings two-fold differences. \textbf{First}, it enables a better formulation of Earth's atmosphere: In every attention block, the Earth-specific positional bias learns different spatial relationship between tokens for different latitudes and heights, hence correcting the non-uniformity brought by the uneven spatial distribution. \textbf{Second}, compared to the original version where all grids share the same bias, the number of learnable parameters of each transformer layer is largely increased from $(2W_\mathrm{pl}-1)\times2(W_\mathrm{lat}-1)\times(2W_\mathrm{lon}-1)$ to $M_\mathrm{pl}\times M_\mathrm{lat}\times W_\mathrm{pl}^2\times W_\mathrm{lat}^2\times(2W_\mathrm{lon}-1)$. In the first block, the latter quantity is about $527\times$ larger than the former one. The huge amount of bias parameters allows each block to flexibly learn specific patterns for each variable, such as the relationship shown in Figure~\ref{fig:irregular_grid}. In practice, we do not observe any difficulties in optimizing the large amount of parameters. Instead, the model converges faster in the training process since useful priors have been introduced. In addition, the Earth-specific positional bias does not increase the FLOPs of the model.

\textbf{Design choices.}\quad We briefly discuss other design choices. Due to the large computational overhead, we do not perform exhaustive ablative or diagnostic studies on the hyper-parameters and we believe there exist configurations that lead to higher accuracy. \textbf{First}, we use $8$ ($2+6$) encoder and decoder layers, which is significantly fewer than the standard Swin transformer. This is to reduce the complexity in both time and memory. If one has a larger GPU memory and a more powerful cluster, increasing the network depth can lead to higher accuracy. \textbf{Second}, it is possible to reduce the number of parameters used in the Earth-specific positional bias by parameter sharing or other techniques. However, we do not consider it as a key issue, because it is unlikely to deploy the weather forecast model to edge device with limited storage. \textbf{Third}, it is possible and promising to feed the weather states of more time points into the model, which changes all tensors from 3D to 4D. While we believe such a modification can lead to accuracy gain, the limited computational budget prevents us from this trial.

\subsection{Hierarchical Temporal Aggregation}
\label{approach:temporal}

\begin{figure}
\centering
\includegraphics[width=8cm]{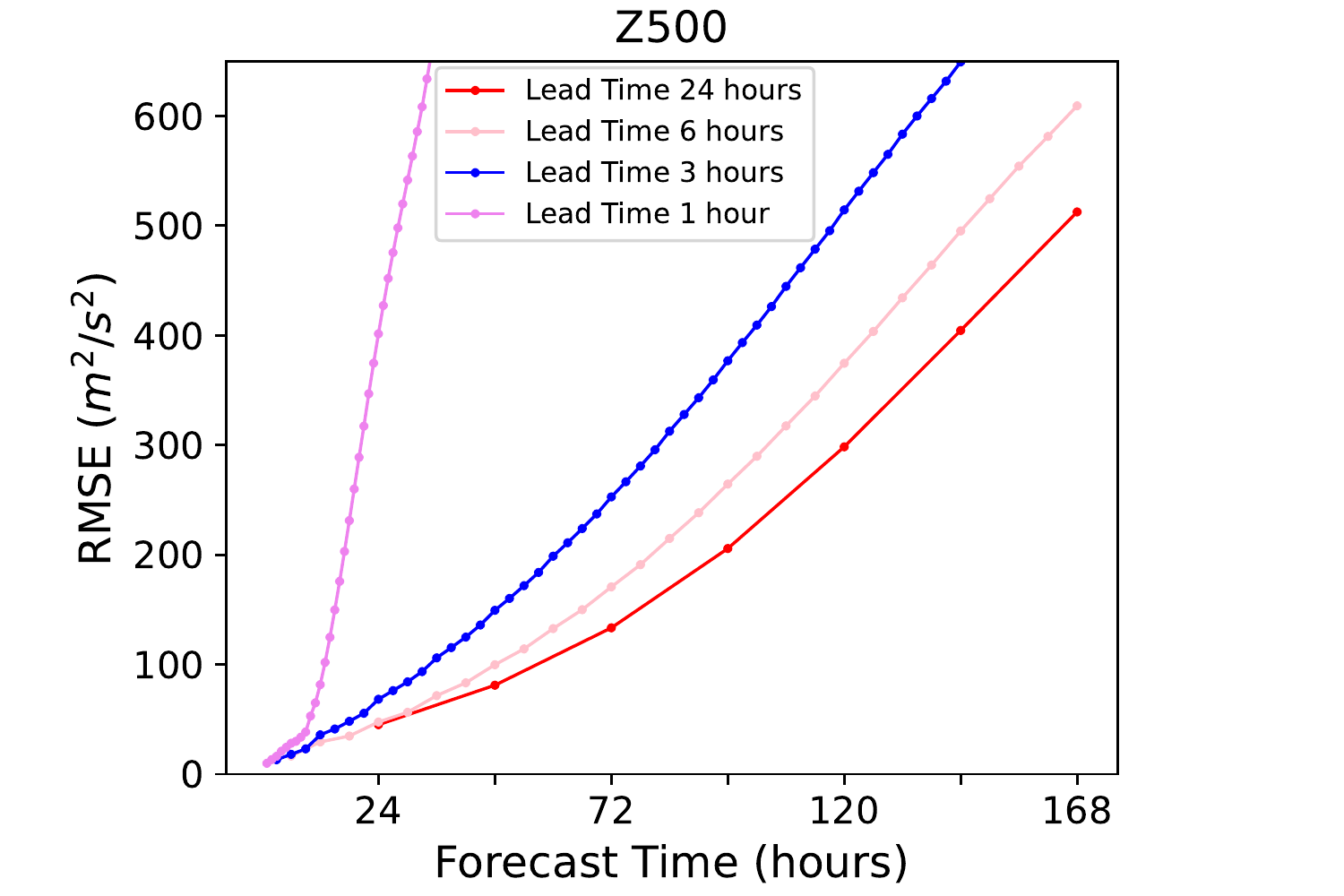}
\caption{The curves showing cumulative forecast errors when one performs up to $7$-day forecast with the base lead time being $1$ hour, $3$ hour, $6$ hours, and $24$ hours, respectively. The statistics are performed in the March 2018 subset.}
\label{fig:leadtime}
\end{figure}

When the goal is to make medium-range weather forecast (\textit{e.g.}, the forecast time is up to $5$ days) yet the lead time of the basic forecast model is relatively short (\textit{e.g.}, FourCastNet trained a model with a lead time of $6$ hours), the system must execute the model many times iteratively, and the cumulative forecast errors can grow continuously. As shown in Figure~\ref{fig:leadtime}, we mimic FourCastNet~\cite{fourcastnet} to execute the $6$-hour model $28$ times to achieve up to $7$-day forecast, and we find that the forecast accuracy rapidly goes down as iteration goes on. Not surprisingly, the forecast accuracy drop becomes dramatic if the basic lead time is set to be $1$ hour (\textit{i.e.}, the model is executed $168$ times), yet the drop is largely alleviated if the lead time is $24$ hours (\textit{i.e.}, executed $7$ times). This implies that, for medium-range and even long-range forecast, the system can benefit much from suppressing the cumulative forecast errors.

For this purpose, we exploit a straightforward yet effective strategy named hierarchical temporal aggregation. We train four individual models for $1$-hour, $3$-hour, $6$-hour, and $24$-hour prediction, respectively. We do not continue enlarging the lead time, because it largely increases the difficulty of training the base model\footnote{We find that, based on the current deep network, it is difficult to perform long-term (say, $28$-day) forecast. We conjecture that, if more powerful methods are used (\textit{e.g.}, using time-aware inputs, increasing the computational complexity, \textit{etc.}), the model may gain such abilities.}. At the testing stage, given a forecast goal, we use the greedy algorithm to guarantee the minimal number of iterations. For example, for $7$-day forecast, we execute $24$-hour forecast $7$ times, while for a $23$-hour forecast, we execute $6$-hour forecast $3$ times, followed by $3$-hour forecast $1$ time and $1$-hour forecast $2$ times.

We point out that hierarchical temporal aggregation makes both the training and testing stages more efficient. For training, it avoids performing recursive optimization as many existing works~\cite{weyn_old,weyn_new,fourcastnet} did, \textit{e.g.}, FourCastNet~\cite{fourcastnet} computed both $f(\mathbf{A})$ and $f(f(\mathbf{A}))$ and produced two loss terms -- although the iterative errors are indeed suppressed, it requires $2\times$ GPU memory for the same model and thus reduced the model size which is one of the critical factors of improvement. In addition, it avoids training a recursive neural network which may be unstable. For testing, especially when the forecast range is large, it reduces the number of forecasts as well as the time complexity.

The four individual models are trained for $100$ epochs using the Adam optimizer. Each full training procedure takes $16$ days on $192$ NVIDIA Tesla-V100 GPUs. We find that all models have not yet arrived at full convergence at the end of $100$ epochs, but the limited computational budget prevents us from continuing the training procedure. A weight decay of $3\times10^{-6}$ and a scheduled DropPath with a drop ratio of $0.2$ are adopted to avoid over-fitting.

\section{Results}
\label{results}

We report the forecast results of Pangu-Weather on two datasets. The first one is the held-out part of ERA5 for an overall evaluation of global, deterministic weather forecast. The second one is the 4th version of International Best Track Archive for Climate Stewardship (IBTrACS) dataset for evaluating the ability at tracking tropical cyclones, a special case of extreme weather forecast.

We compare Pangu-Weather to the strongest methods in both worlds of NWP and AI, namely, operational IFS offered by ECMWF (downloaded from the TIGGE archive~\cite{tigger_data})\footnote{We failed to download part of forecast results of the surface variables from TIGGE, due to the unavailability of ECMWF's Data Handling Systems from September to November, so we compare our results to operational IFS by (i) fetching the numbers reported in WeatherBench~\cite{weatherbench} and (ii) extracting the quantities from the plots in the FourCastNet paper~\cite{fourcastnet}.} and FourCastNet~\cite{fourcastnet}. For tropical cyclones tracking, we also download ECMWF-HRES forecast as a stronger competitor against Pangu-Weather in IBTrACS. To the best of our knowledge, no prior AI-based methods have ever reported quantitative results for tropical cyclones tracking.

\begin{figure*}
\centering
\includegraphics[width=17cm]{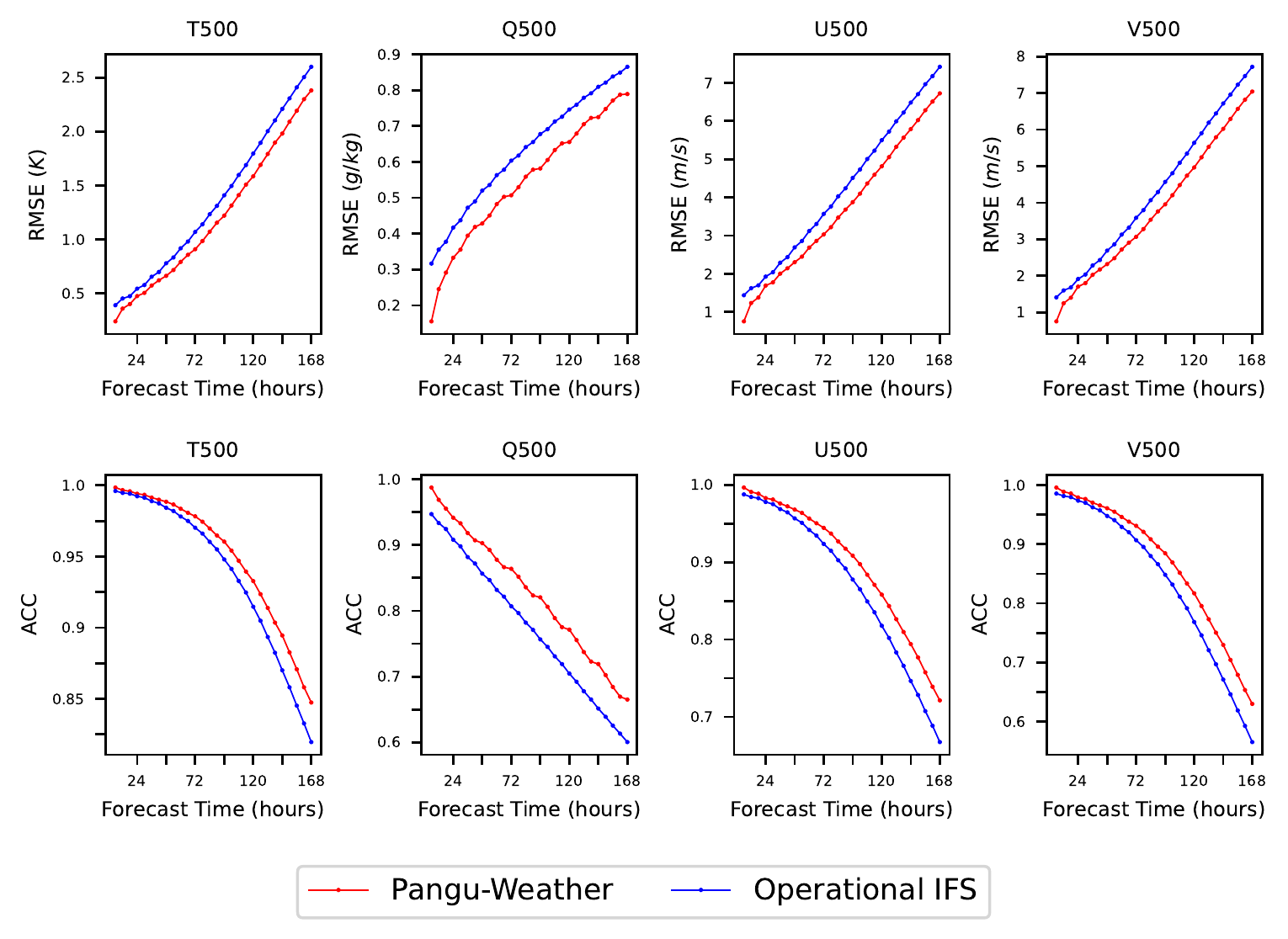}
\caption{The comparison of forecast accuracy in terms of latitude-weighted RMSE (lower is better) and ACC (higher is better) of four upper-air variables at the pressure level of $500\mathrm{hPa}$. Here, T, Q, U and V stand for temperature, specific humidity, $u$-component and $v$-component of wind speed, respectively.}
\label{fig:upperair}
\end{figure*}

\subsection{Deterministic Forecast}
\label{results:deterministic}

The deterministic forecast of Pangu-Weather is performed on the unperturbed initial states from ERA5. The forecast resolution of Pangu-Weather is determined by the training data (\textit{i.e.}, ERA5), where the spatial resolution is $0.25^\circ\times0.25^\circ$, comparable to the control forecast of ECMWF ENS product~\cite{ecmwf_method} and same as FourCastNet~\cite{fourcastnet}, yet the spacing of forecast (the minimal forecast time) is $1$ hour (\textit{i.e.}, Pangu-Weather can provide hour-by-hour forecast), $6\times$ smaller than that of FourCastNet~\cite{fourcastnet}.

Following the prior AI-based methods, the accuracy of deterministic forecast is computed by two quantitative metrics, namely, the latitude-weighted Root Mean Square Error (RMSE) and latitude-weighted Anomaly Correlation Coefficient (ACC). For a specified time point $t$, the RMSE and ACC of any variable $v$ (\textit{e.g.}, 2m temperature or $500\mathrm{hPa}$ geopotential) are defined as follows:
\begin{equation}
\label{eqn:rmse}
\mathrm{RMSE}(v,t)=\sqrt{\frac{\sum_{i=1}^{N_\mathrm{lat}}\sum_{j=1}^{N_\mathrm{lon}}{L(i)(\hat{\mathbf{A}}_{i,j,t}^v-\mathbf{A}_{i,j,t}^v)^2}}{N_\mathrm{lat}\times N_\mathrm{lon}}},
\end{equation}
\begin{equation}
\label{eqn:acc}
\mathrm{ACC}(v,t)=\frac{\sum_{i,j}{L(i)}\hat{\mathbf{A}}_{i,j,t}^{\prime v}\mathbf{A}_{i,j,t}^{\prime v}}{\sqrt{\sum_{i,j}{L(i)}(\hat{\mathbf{A}}_{i,j,t}^{\prime v})^2\times\sum_{i,j}{L(i)}(\mathbf{A}_{i,j,t}^{\prime v})^2}},
\end{equation}
where $L(i)=N_\mathrm{lat}\times\frac{\cos\phi_i}{\sum_{i'=1}^{N_\mathrm{lat}}{\cos\phi_{i'}}}$ stands for the weight at latitude $\phi_i$ and $\mathbf{A}'$ denotes the difference between $\mathbf{A}$ and the climatology (\textit{i.e.}, long-term mean of weather states, which is estimated on the training data over $39$ years). Note that we omitted the range of summation in Eqn~\eqref{eqn:acc} for simplicity. In what follows, we report these two metrics on upper-air atmospheric variables and surface weather variables to show the superiority of Pangu-Weather. We also provide extensive visualization and diagnostic results for qualitative studies.

\begin{figure*}
\centering
\includegraphics[width=17cm]{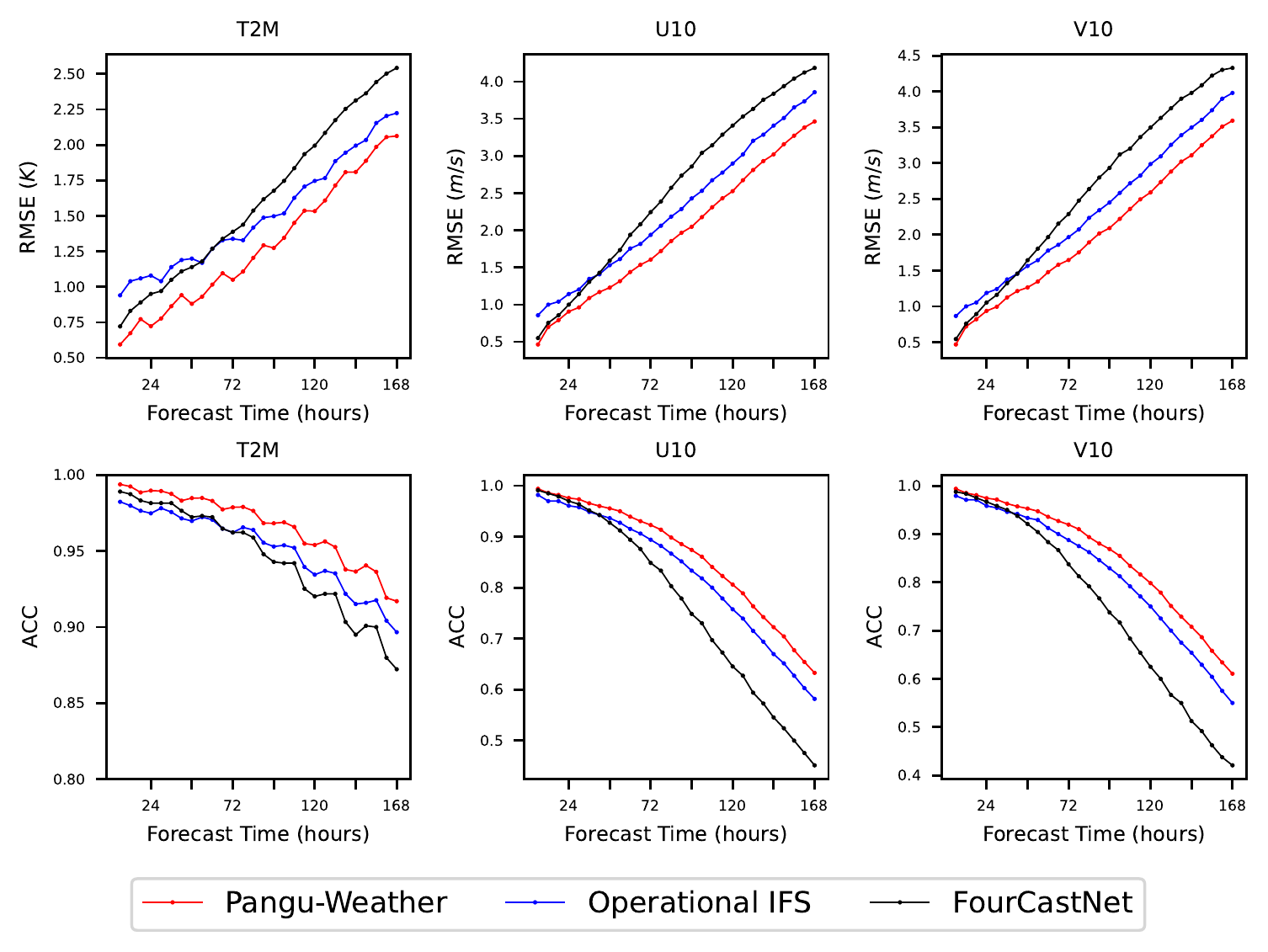}
\caption{The comparison of forecast accuracy in terms of latitude-weighted RMSE (lower is the better) and ACC (higher is the better) of three surface variables. Here, T2M, U10, V10 stand for 2m temperature, $u$-component and $v$-component of 10m wind speed, respectively.}
\label{fig:surface_general_results}
\end{figure*}

\subsubsection{Upper-air Atmospheric Variables}
\label{results:deterministic:upperair}

As in the training procedure (see Section~\ref{approach:data} for data preparation), Pangu-Weather forecasts five important upper-air variables (\textit{i.e.}, geopotential, specific humidity, temperature, $u$-component and $v$-component of wind speed) at $13$ pressure levels (\textit{i.e.}, $50\mathrm{hPa}$, $100\mathrm{hPa}$, $150\mathrm{hPa}$, $200\mathrm{hPa}$, $250\mathrm{hPa}$, $300\mathrm{hPa}$, $400\mathrm{hPa}$, $500\mathrm{hPa}$, $600\mathrm{hPa}$, $700\mathrm{hPa}$, $850\mathrm{hPa}$, $925\mathrm{hPa}$, and $1000\mathrm{hPa}$), with a spatial resolution of $0.25^\circ\times0.25^\circ$. This is to maximally ease the comparison to operational IFS~\cite{tigger_data} and FourCastNet~\cite{fourcastnet}, the best NWP and AI-based methods.

The testing environment is established on the weather data in 2018\footnote{We also test our system in the 2020 and 2021 data, while we cannot provide comparative results since no prior works have reported results on these data. The property of forecast results is mostly similar to that observed in the 2018 data.}. Following the protocol of operational IFS, we choose $2$ time points (00:00 UTC and 12:00 UTC) each day as the initial time\footnote{The test points on Jan 1st, 2018 are excluded due to the overlap with training data. All test points in December 2018 are unavailable due to a server error of ECMWF. In addition, for the T850 variable, all test points in October 2018 are not used due to an unexpected error of data download from the TIGGE archive.} and produce hourly forecast for the upcoming week, namely, forecast time being $1\mathrm{h},2\mathrm{h},\ldots,168\mathrm{h}=7\mathrm{d}$. Quantitative comparisons are mainly made between Pangu-Weather and operational IFS, while the comparisons against other AI-based methods (\textit{e.g.}, \cite{weatherbench,weyn_old,weyn_new,swinvrnn,fourcastnet,graph_weather}) are incomplete due to the difference in spatial resolutions, test subsets, post-processing methods, \textit{etc}. Note that all prior AI-based methods reported inferior forecast accuracy compared to operational IFS, while our method significantly outperforms operational IFS, claiming clear advantages over these candidates. For two specific variables, Z500 and T850, we directly compare Pangu-Weather against FourCastNet by fetching the numerical values from the plots in the paper -- this introduces some errors which are negligible compared to the accuracy gap between Pangu-Weather and FourCastNet.

\begin{figure*}
\centering
\includegraphics[width=9cm]{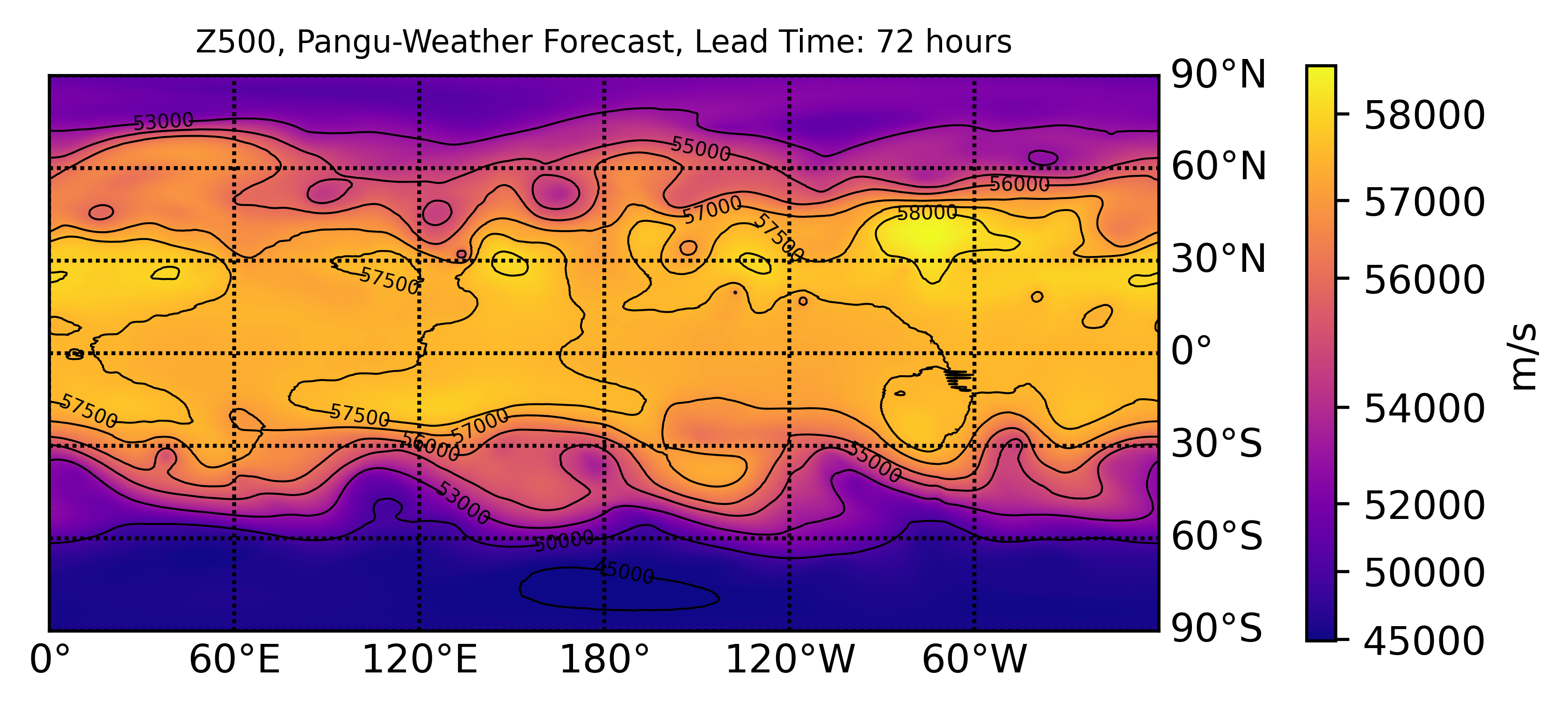}\hfill
\includegraphics[width=9cm]{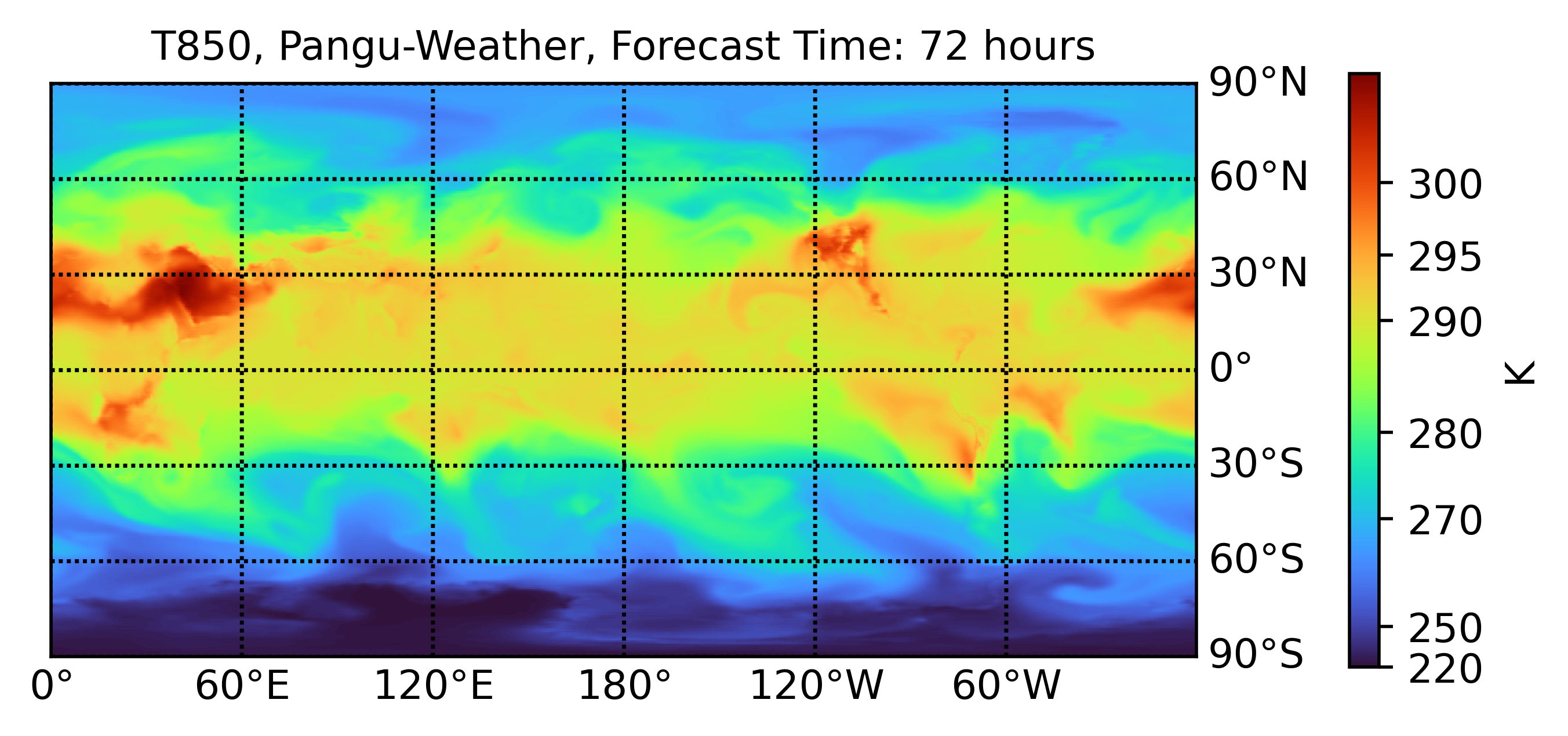}\\
\includegraphics[width=9cm]{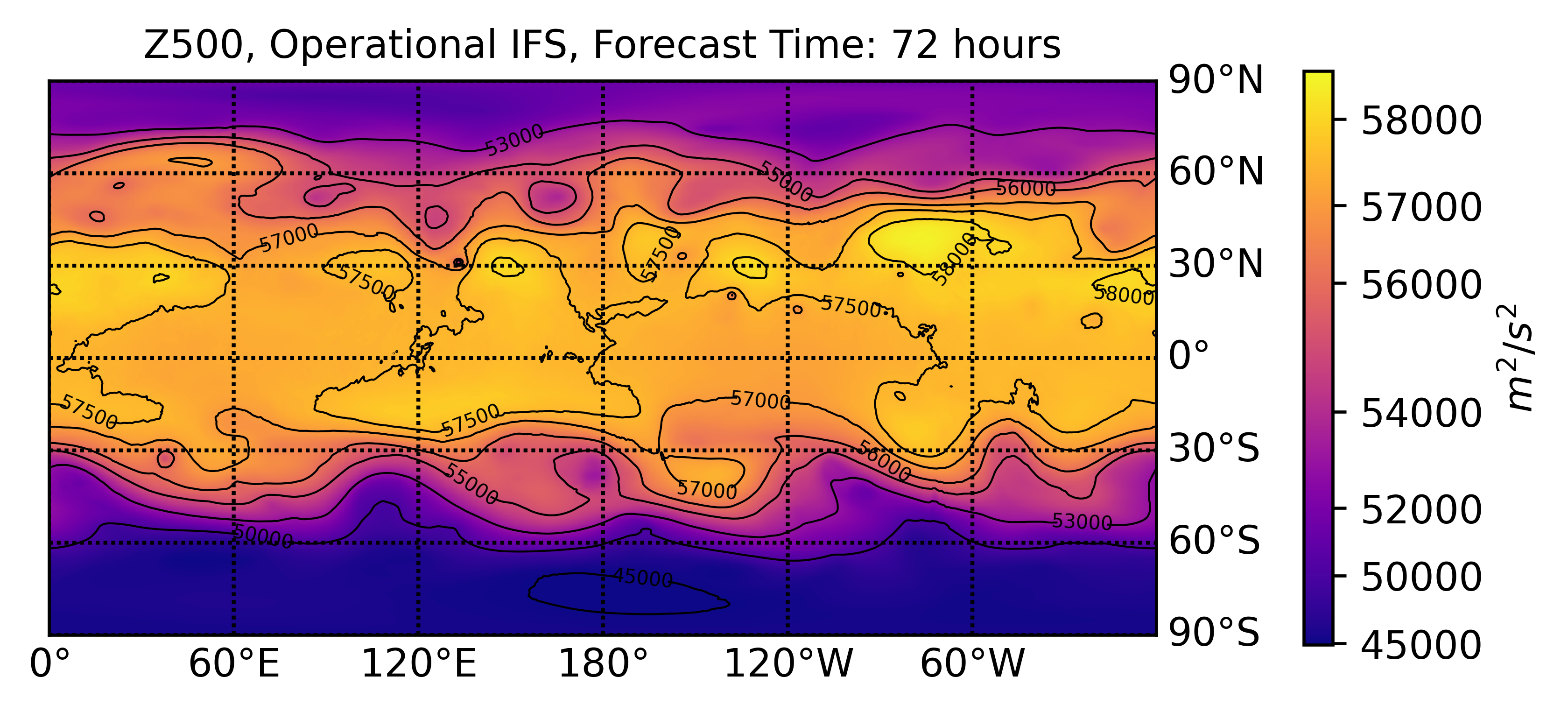}\hfill
\includegraphics[width=9cm]{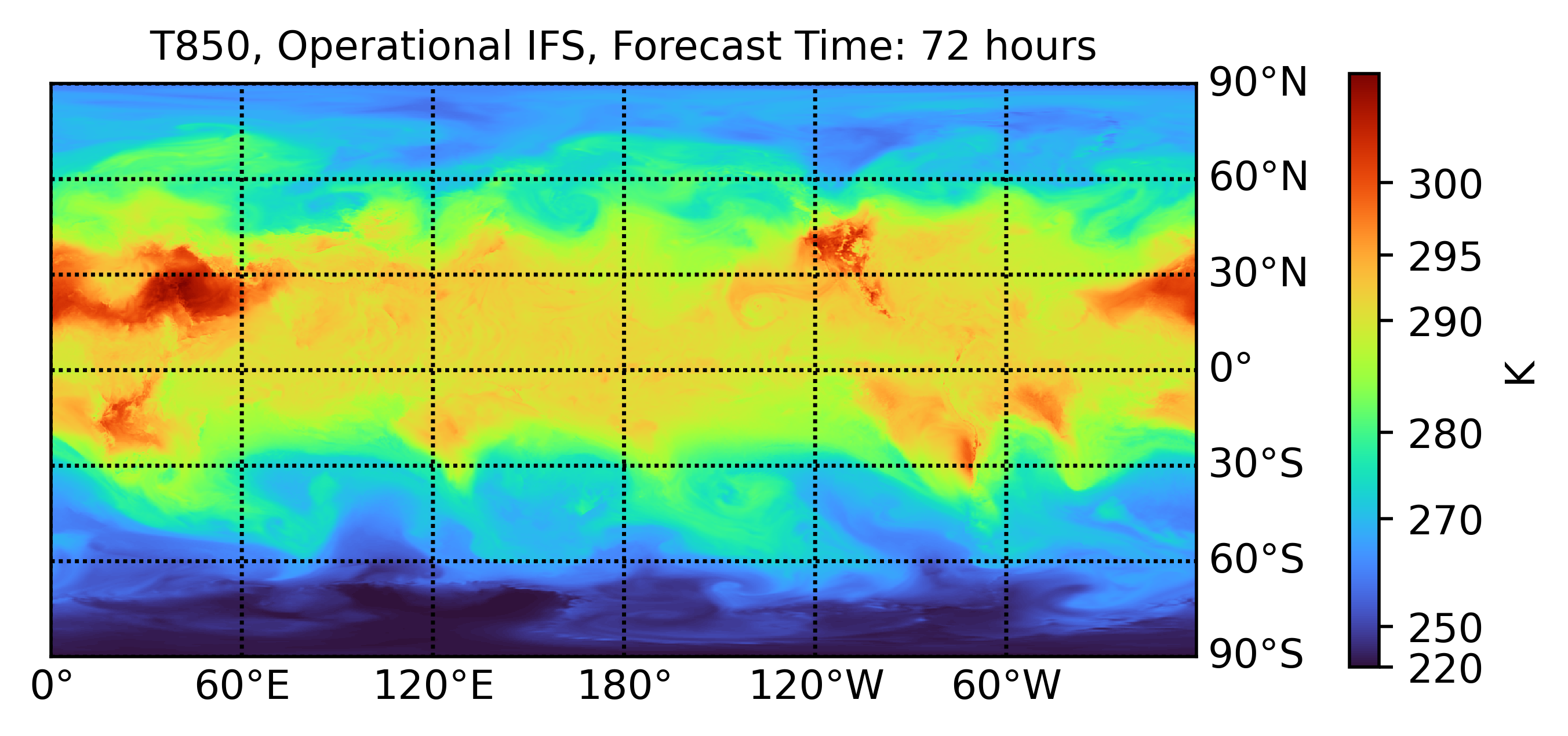}\\
\includegraphics[width=9cm]{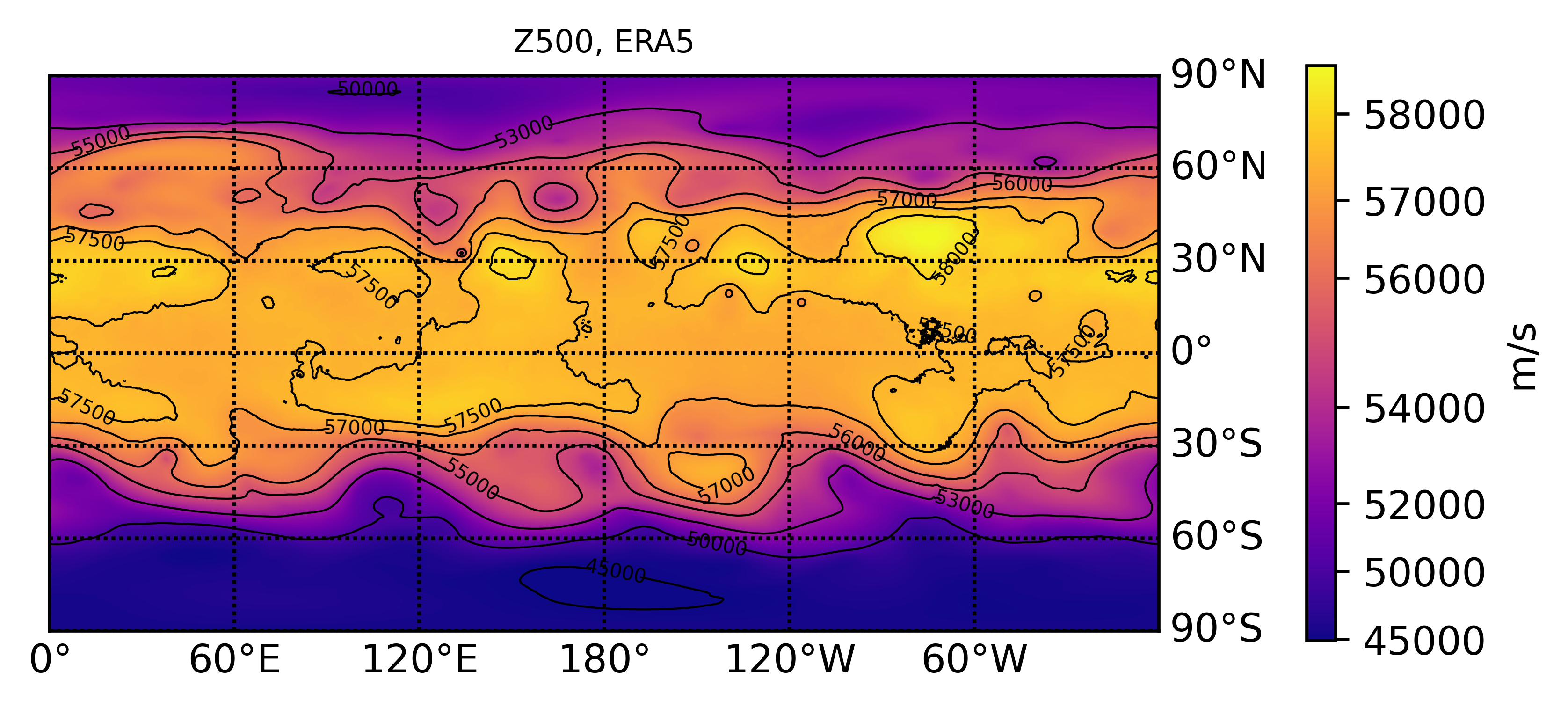}\hfill
\includegraphics[width=9cm]{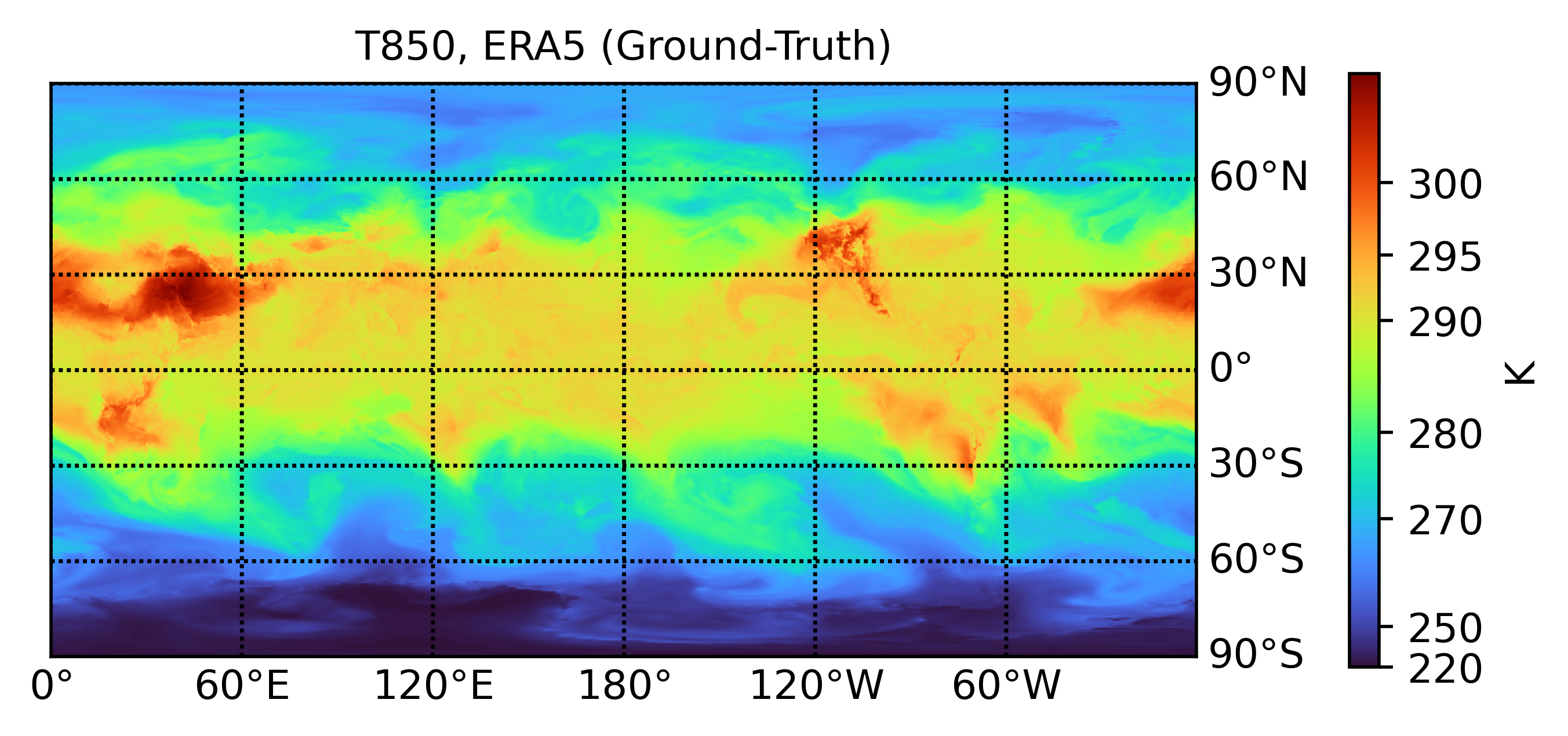}
\caption{Visualization of $3$-day weather forecast produced by Pangu-Weather (top), operational IFS (middle), and the ERA5 ground-truth (bottom). The left and right columns show the maps of $500\mathrm{hPa}$ geopotential (Z500) and $850\mathrm{hPa}$ temperature (T850), respectively. The input time point (\textit{i.e.} the forecast is performed on) is 00:00 UTC, September 1st, 2018.}
\label{fig:upper_visualize}
\end{figure*}

The comparative results between Pangu-Weather and operational IFS are shown in Figure~\ref{fig:overview} (top) and Figure~\ref{fig:upperair}, where Pangu-Weather enjoys consistent gains (in \textbf{all} forecast times and for all variables) in forecast accuracy compared to operational IFS. The advantage becomes more significant as forecast time increases, implying that AI-based methods are better at capturing effective (though non-interpretable) patterns for medium-range weather forecast. Specifically, we note that the `forecast time gain' of Pangu-Weather over operational IFS (\textit{i.e.}, the difference between forecast times at the same forecast accuracy) is more than $12$ hours for all variables and more than $24$ hours for specific humidity -- this implies that AI-based methods are significantly better at forecasting specific variables.

Specifically, we investigate $500\mathrm{hPa}$ geopotential (Z500) and $850\mathrm{hPa}$ temperature (T850), the variables that were widely reported in prior AI-based methods. The quantitative comparison for these two variables is shown in Figure~\ref{fig:overview}. As shown, the forecast accuracy of Pangu-Weather is consistently higher than that of operational IFS and FourCastNet, the previous best AI-based method (yet weaker than operational IFS). Quantitatively, for Z500, the $3$-day and $5$-day RMSEs (in $\mathrm{m}^2/\mathrm{s}^2$) of operational IFS are $152.8$ and $333.7$, respectively, and Pangu-Weather reduces them to $134.5$ and $296.7$ ($133.9$ and $294$ if the December 2018 data are included). For T850, the $3$-day and $5$-day RMSEs (in $\mathrm{K}$) of operational IFS are $1.37$ and $2.06$, respectively, and Pangu-Weather reduces them to $1.14$ and $1.79$ ($1.13$ and $1.77$ if the December 2018 data are included), claiming an over $10\%$ relative error drop. The relative drop of RMSE is more than $10\%$ in all scenarios, which also reflects in a `forecast time gain' of $10$--$15$ hours. When compared to FourCastNet, we observe even more significant accuracy gains -- the relative reduction of RMSE is more than $30\%$ in the above scenarios, and the `forecast time gain' is also enlarged to more than $36$ hours.

\subsubsection{Surface Weather Variables}
\label{results:deterministic:surface}

Pangu-Weather forecasts four important surface variables, \textit{i.e.}, 2m temperature, $u$-component and $v$-component of 10m wind speed, and mean sea level pressure. Compared to the upper-air variables, these surface variables have close and complex relationship to topography and human activities (\textit{e.g.}, urban heat island effect) and thus are more difficult to forecast. The testing environment is established in a similar way of forecasting upper-air variables\footnote{The test points on Jan 1st, 2018 are excluded due to the overlap with training data.}. Quantitative comparisons are made between Pangu-Weather and previous best NWP method (\textit{i.e.}, operational IFS) and AI-based method (\textit{i.e.}, FourCastNet~\cite{fourcastnet}), where the numerical results of FourCastNet and operational IFS are fetched from the plots in the paper\footnote{For a fair comparison to FourCastNet, we follow the protocol to set the test interval (the gap between neighborhood test time points) to be $9$ days for T2M and $2$ days for U10 and V10, albeit we can produce forecast results every single hour. We also report the RMSE values using a fixed $6$-hour interval in the following part.}. The comparative results are shown in Figure~\ref{fig:surface_general_results}. Again, Pangu-Weather outperforms both competitors in terms of forecast accuracy in \textbf{all} forecast times, for all variables, and the advantage becomes more significant as forecast time increases. The `forecast time gain' of all these variables is about $18$ hours, slightly longer than the gain in forecasting upper-air variables.

We investigate the forecast accuracy of separate variables. For 2m temperature (T2M), the $3$-day and $5$-day RMSEs (in $\mathrm{K}$) are $1.34$ and $1.75$ for operational IFS, $1.39$ and $2.00$ for FourCastNet, and Pangu-Weather reduces them to $1.05$ and $1.53$ ($1.06$ and $1.52$ with a $6$-hour test interval), respectively. For $u$-component of 10m wind speed (U10), the $3$-day and $5$-day RMSEs (in $\mathrm{m}/\mathrm{s}$) are $1.94$ and $2.90$ for
operational IFS, $2.24$ and $3.41$ for FourCastNet, and Pangu-Weather reduces them to $1.61$ and $2.53$ ($1.61$ and $2.55$ for the test data in 2018 with a $6$-hour interval), respectively. We omit the numerical comparison for $v$-component of 10m wind speed (V10) since it is almost the same as that of U10.

We have the forecast results for the variable of mean sea level pressure (MSLP) but we cannot provide the quantitative comparison due to the unavailability of data from the ECMWF server. We believe that our forecast is the best candidate because of two reasons. On the one hand, According to prior experiences~\cite{ecmwf_2009verification}, a model that better forecasts on other surface variables (\textit{i.e.}, T2M, U10, V10) also enjoys a higher forecast accuracy on MSLP. On the other hand, we make use of our forecast of MSLP for tracking tropical cyclones -- as shown in Section~\ref{results:extreme:cyclones}, Pangu-Weather achieves much better results, quantitatively and qualitatively, than operational IFS in forecasting $88$ named tropical cyclones in the year of 2018.

In addition, we evaluate Pangu-Weather on WeatherBench~\cite{weatherbench}, a benchmark for low-resolution weather forecast. For this purpose, we simply down-sample the forecast results of Pangu-Weather by $22.5\times$ into a coarse grid with a spatial resolution of $5.625^\circ\times5.625^\circ$, and compare the results to the down-sampled ERA5 ground-truth. Quantitatively, operational IFS reported $3$-day/$5$-day RMSEs of T2M\footnote{We only show the comparison on T2M, because WeatherBench evaluated the forecast results of T850 and Z500 on the 2017 subset which is part of our training data.} being $1.35/1.77$ on WeatherBench, and Pangu-Weather improves the results to $1.04$/$1.51$, respectively.

\begin{figure*}
\centering
\includegraphics[width=17cm]{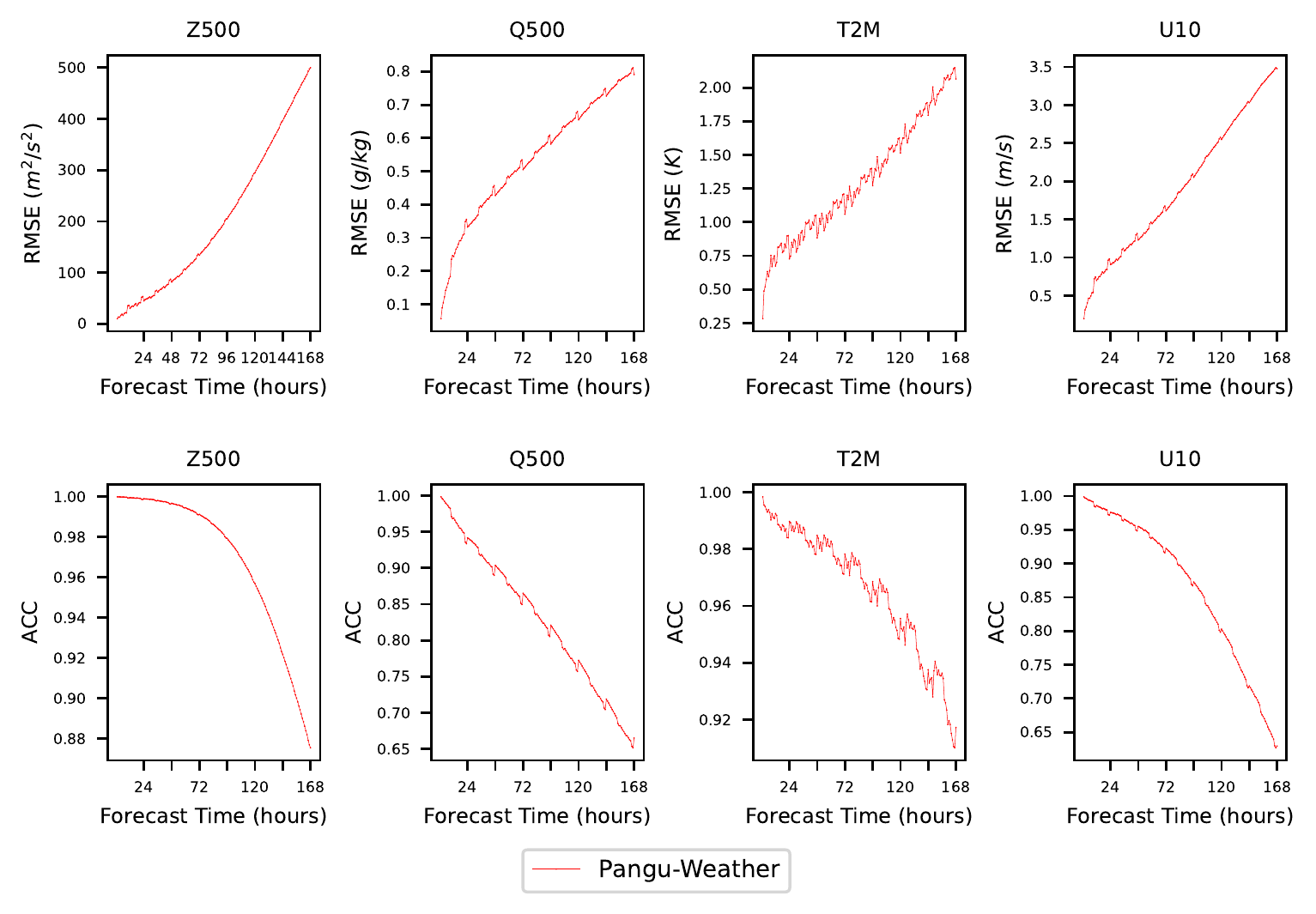}
\caption{Hourly forecast results for two upper-air variables ($500\mathrm{hPa}$ geopotential, Z500, and $500\mathrm{hPa}$ specific humidity, Q500) and two surface variables (2m temperature, T2M, and $u$-component of 10m wind speed, U10). The forecast time ranges from $1$ hour to $7$ days ($168$ hour), and both latitude-weighted RMSE (lower is better) and ACC (higher is better) are reported. The input time points are chosen from the $2018$ data -- since no comparison is made, we only exclude the data on January 1st due to the overlaps with the training set.}
\label{fig:hourly_skill}
\end{figure*}

\subsubsection{Visualization}
\label{results:deterministic:visualization}

In Figure~\ref{fig:upper_visualize}, we first visualize the $72$-hour forecast of Pangu-Weather on two upper-air variables, namely, Z500 and T850, and compare the results to operational IFS and the ERA5 ground-truth. Both forecast results are sufficiently close to the ground-truth, yet one can detect the differences between them. Pangu-Weather produce smoother contour lines, implying that the model tends to forecast similar values for neighboring regions -- this is a typical property of deep neural networks in learning from large-scale datasets. In comparison, operational IFS tends to preserve small-scale structures, yet such predictions are not guaranteed to be correct. As shown in the previous part, Pangu-Weather enjoys the advantage of overall forecast accuracy.


In Figure~\ref{fig:overview} (top left), we also visualize the $72$-hour forecast of Pangu-Weather on two surface variables, namely, 2m temperature (T2M) and 10m wind speed ($\sqrt{u^2+v^2}$). As can be seen, Pangu-Weather produces high-resolution forecasts that are (i) very close to the ERA5 ground-truth (also refer to the previous part for quantitative results) and (ii) sufficient to preserve most of small-scale structures of surface variables.

\begin{figure*}
\centering
\includegraphics[width=17cm]{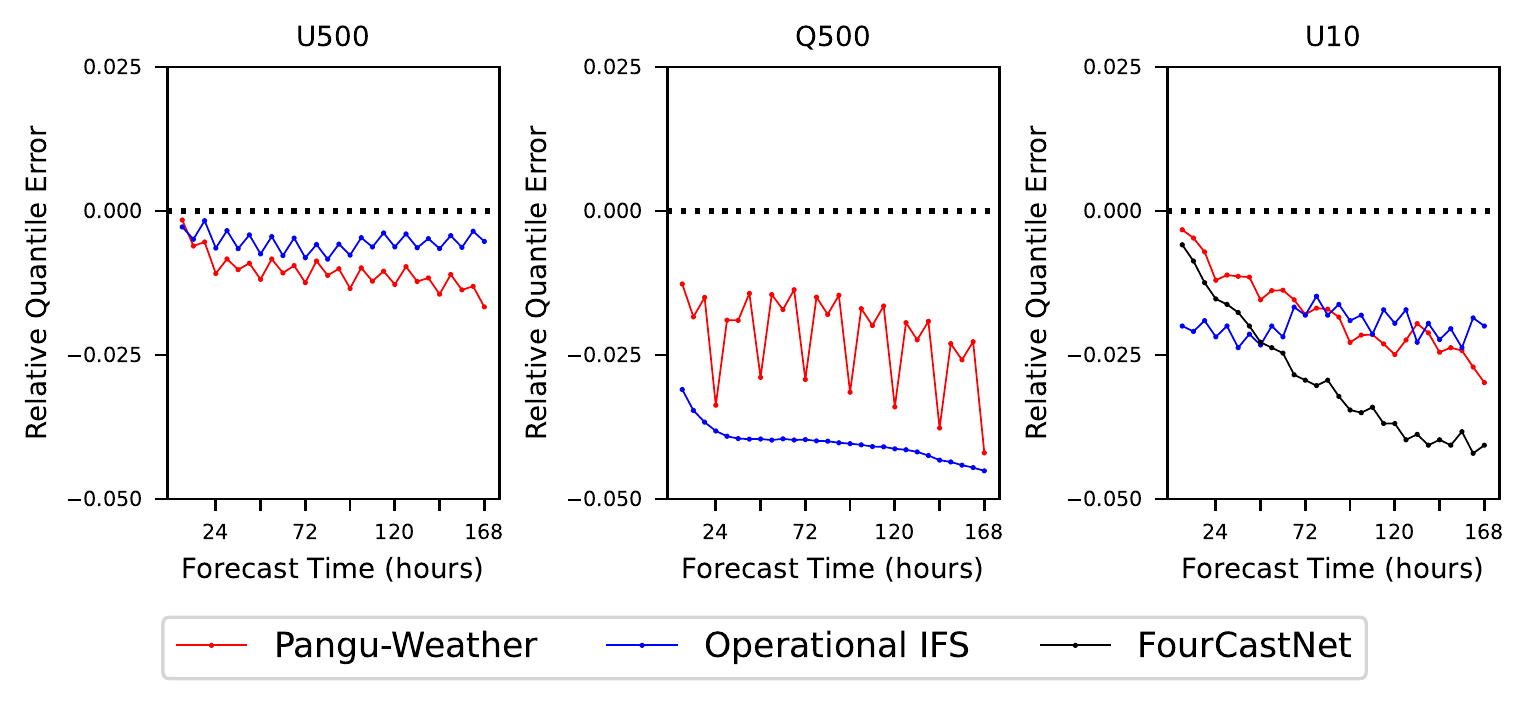}
\caption{Plots of RQE values with respect to forecast time for two upper-air variables ($500\mathrm{hPa}$ $u$-component of wind speed, U500, and $500\mathrm{hPa}$ specific humidity, Q500) and one surface variables ($u$-component of 10m wind speed, U10).}
\label{fig:overall_tendency}
\end{figure*}

\subsubsection{Diagnostic Studies}
\label{results:deterministic:visualization}

We investigate the monthly averaged $5$-day latitude-weighted ACC of four upper-air variables, namely, geopotential (Z), specific humidity (Q), temperature (T), and $u$-component of wind speed (U), all at the pressure level of $500\mathrm{hPa}$. The comparison between Pangu-Weather and operational IFS is shown in Figure~\ref{fig:overview} (top right). Pangu-Weather outperforms operational IFS in every single month, demonstrating the stability of forecast. More importantly, the advantage of Pangu-Weather becomes more significant in the worst performed months (\textit{e.g.}, April and May), implying that AI-based methods have learned useful and complementary knowledge from large data. We conjecture that such knowledge may correspond to (i) unknown or unformulated atmospheric procedures or (ii) better manipulations with missing factors. Studying these factors may be an interesting topic for meteorologists.


A clear advantage of Pangu-Weather lies in its ability of performing hourly weather forecast. We plot the hourly RMSE and ACC values for two upper-air variables (Z500 and Q500) and two surface variables (T2M and U10) in Figure~\ref{fig:hourly_skill}. Note that we have applied a greedy algorithm based on hierarchical temporal aggregation as elaborated in Section~\ref{approach:temporal}. For some variables (\textit{e.g.}, Q500 and T2M), we observe a clear trend that forecast accuracy drops with the number of iterations, \textit{e.g.}, $3$ calls are required for a $72$-hour forecast, while $8$ calls are required for a $71$-hour forecast (\textit{i.e.}, $71=24+24+6+6+6+3+1+1$), and thus the $71$-hour accuracy is much lower due to cumulative forecast errors. While we can easily improve the accuracy by moving the time point back (\textit{e.g.}, performing $72$-hour forecast using $1$-hour-earlier weather states for $71$-hour forecast), we just offer the original forecast results here to show this important phenomenon. This calls for an advanced temporal aggregation algorithm in the future -- one possibility lies in integrating the time axis into input data and making use of 4D deep neural networks, but this implies much heavier computational overheads.

\subsubsection{Computational Costs}

A clear advantage of AI-based methods lies in the inference speed. FourCastNet~\cite{fourcastnet} claimed a $45\rm{,}000\times$ speedup over the traditional NWP method, and Pangu-Weather is comparable with FourCastNet (see the next paragraph). Considering the advantage in forecast accuracy, Pangu-Weather has the potentials of replacing conventional NWP, enabling real-time weather forecast to be performed any time (\textit{e.g.}, once a second), rather than the current status that weather forecast is performed merely a few times per day. A few side benefits are expected. (i) It largely increases the timeliness of short-range weather forecast which is important in warning about short-term extreme weathers, \textit{e.g.}, cloudbursts. (ii) It enables large-member ensemble forecast which is important for meteorologists to pay attentions to the sensitive weather factors or variables.

The inference speed of Pangu-Weather is comparable to that of FourCastNet~\cite{fourcastnet}, implying that using holistic 3D deep neural networks for inference is slightly more costly than using 2D counterparts, yet the accuracy is much higher. In a system-level comparison, FourCastNet requires $280\mathrm{ms}$ for inferring a $24$-hour forecast on an NVIDIA Tesla-A100 GPU ($312$ TeraFLOPS), while Pangu-Weather needs $1\rm{,}400\mathrm{ms}$ on an NVIDIA Tesla-V100 GPU ($120$ TeraFLOPS). Taking GPU performance into consideration, Pangu-Weather is about $50\%$ slower than FourCastNet, while still being one of the fastest systems for high-resolution, global weather forecast.

To further accelerate Pangu-Weather, we can train models with larger lead times (\textit{e.g.}, $72$ hours) so as to reduce the number of temporal aggregations. We expect to explore this direction in the near future.

\subsection{Results on Extreme Weather Events}
\label{results:extreme}

Extreme weather forecast plays a vital role of global weather forecast. Despite rare occurrence, extreme weather events like hurricanes can bring tremendous casualty and economical loss. Therefore, it is expected that weather forecast systems can warn about upcoming extreme weather events that complement daily weather reports.

In this subsection, we investigate the ability of Pangu-Weather in forecasting extreme weather events and compare the ability to that of conventional NWP methods, \textit{i.e.}, operational IFS. We first introduce a quantitative metric named relative quantile error to measure the overall tendency in Section~\ref{results:extreme:tendency}, and then study a special and important case, \textit{i.e.}, tracking tropical cyclones, in Section~\ref{results:extreme:cyclones}.

\subsubsection{Overall Tendency in Predictions of Extremes}
\label{results:extreme:tendency}

We use a similar approach to~\cite{fildier2021distortions} to compare the values of top-level quantiles calculated on the forecast result and ground-truth. Mathematically, we set $D=50$ percentiles, denoted as $q_1,\ldots,q_D$. Following FourCastNet~\cite{fourcastnet}, we set $q_1=90\%$, $q_D=99.99\%$, and the intermediate ones are linearly distributed between $q_1$ and $q_D$ in the logarithmic scale. Then, the corresponding quantiles, denoted as $Q_1,\ldots,Q_D$, are computed individually for each pair of weather variable and forecast time, \textit{e.g.}, for all $3$-day forecasts of U10, pixel-wise values are summarized from all frames for statistics. Finally, the relative quantile error (RQE) is used for measuring the difference between the ground-truth and any weather forecast system:
\begin{equation}
\mathrm{RQE}=\sum_{d=1}^D\frac{\hat{Q}_d-Q_d}{Q_d},
\end{equation}
where $Q_d$ and $\hat{Q}_d$ are different versions of the $d$-th quantile calculated on the ERA5 ground-truth and the system being investigated, \textit{e.g.}, Pangu-Weather. RQE can measure the overall tendency, where $\mathrm{RQE}<0$ and $\mathrm{RQE}>0$ imply that the forecast system tends to underestimate and overestimate the intensity of extremes, respectively.

In Figure~\ref{fig:overall_tendency}, we plot the RQE values for two upper-air variables (U500, V500) and one surface variable (U10) with respect to forecast time. Pangu-Weather is compared to both operational IFS and FourCastNet (only for U10). As seen, all the three methods tend to underestimate extremes, \textit{i.e.}, the RQE values are consistently smaller than $0$. The absolute RQE values reported by AI-based methods generally grow (\textit{i.e.}, heavier underestimation) with forecast time, while that of operational IFS remains mostly unchanged. We attribute the above observation to the cumulative forecast errors of AI-based methods -- compared to FourCastNet, Pangu-Weather significantly alleviates such errors with hierarchical temporal aggregation (see Section~\ref{approach:temporal}). Compared to operational IFS, Pangu-Weather shows higher absolute RQE values (\textit{i.e.}, heavier underestimation) for U500 and lower absolute RQE values (\textit{i.e.}, lighter underestimation) for Q500. Regarding U10, Pangu-Weather is much better than operational IFS for up to $3$ days ($72$ hours) and then becomes slightly worse due to cumulative forecast errors.

\begin{figure}
\centering
\includegraphics[width=6.6cm]{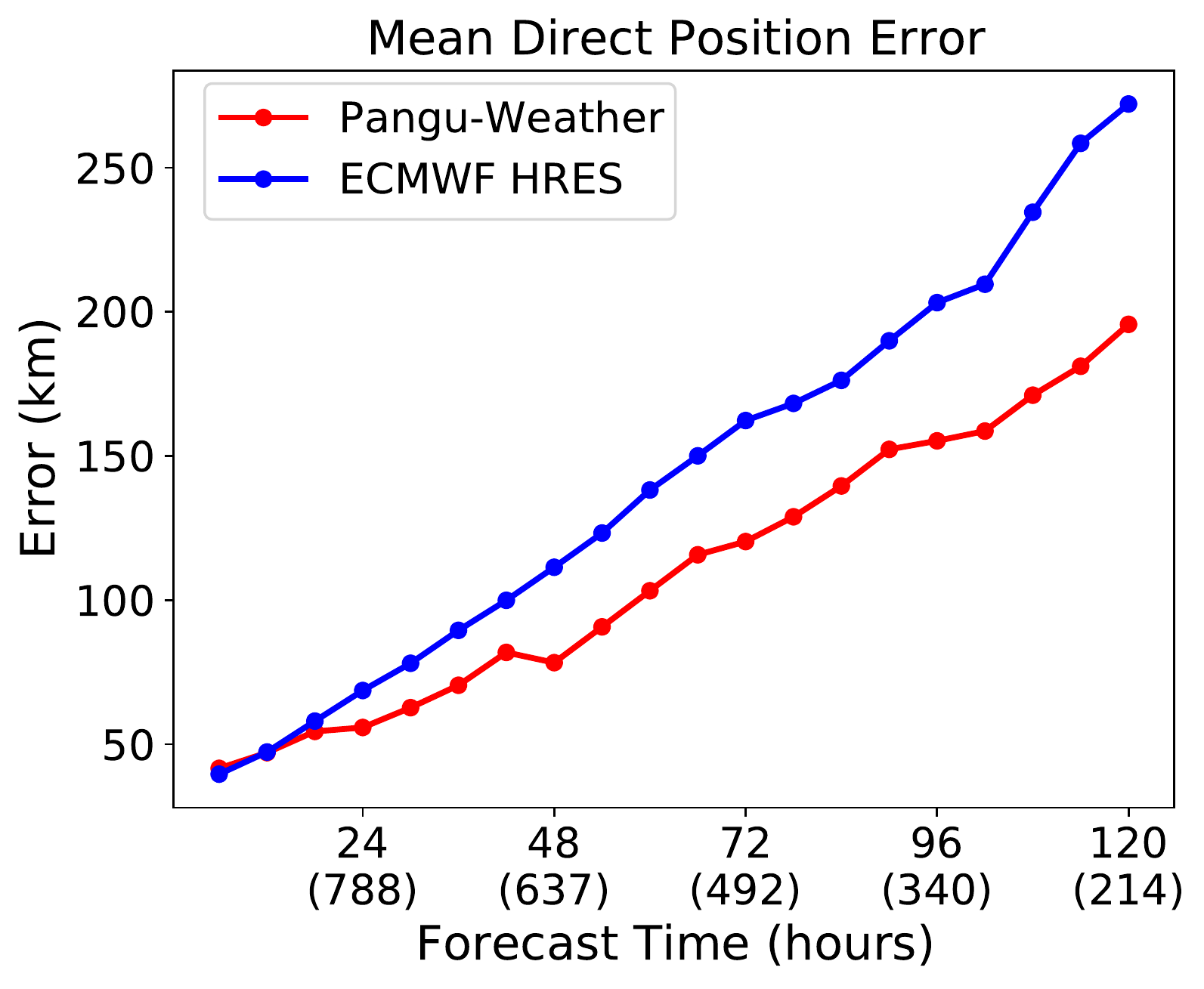}
\includegraphics[width=8cm]{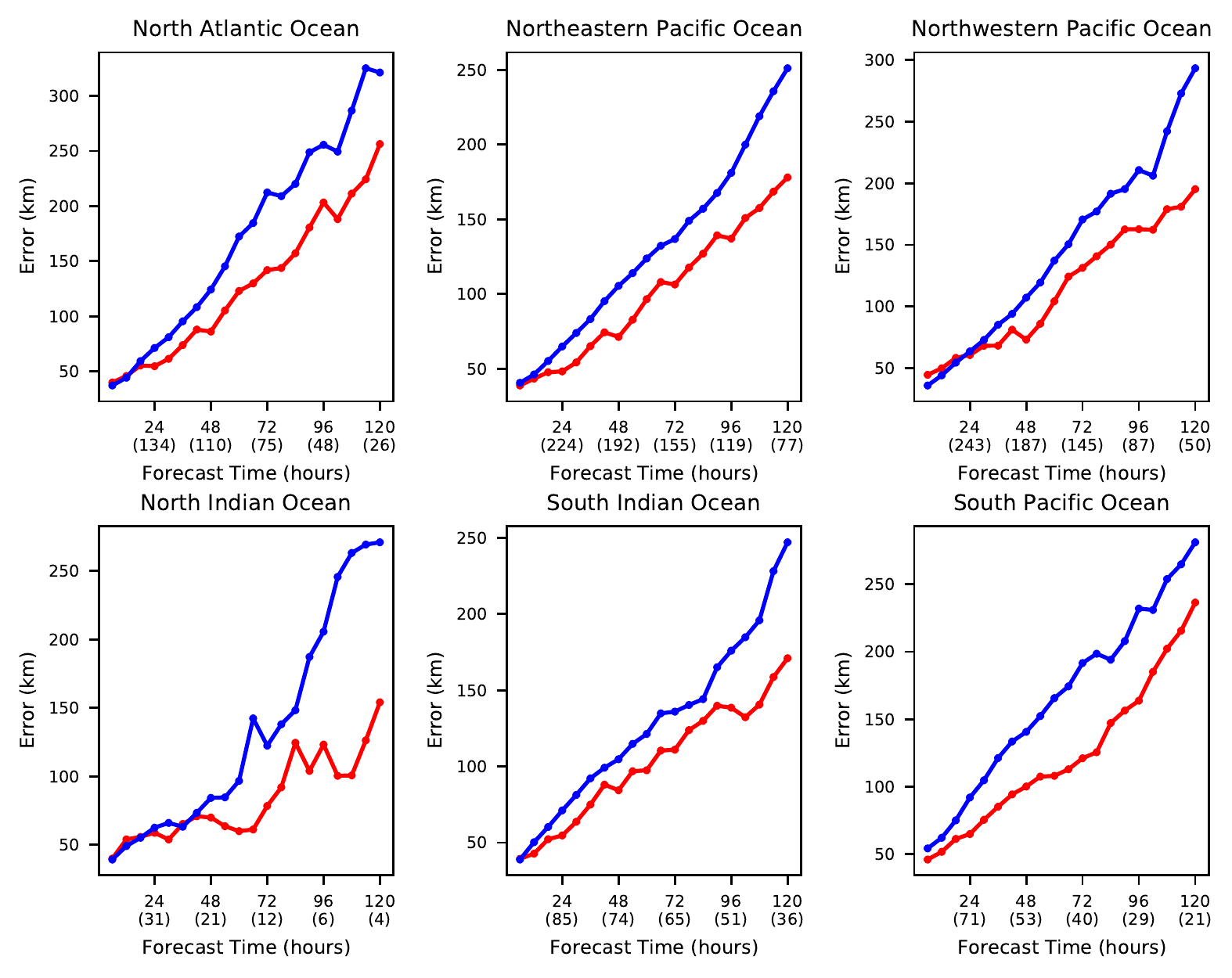}
\includegraphics[width=8cm]{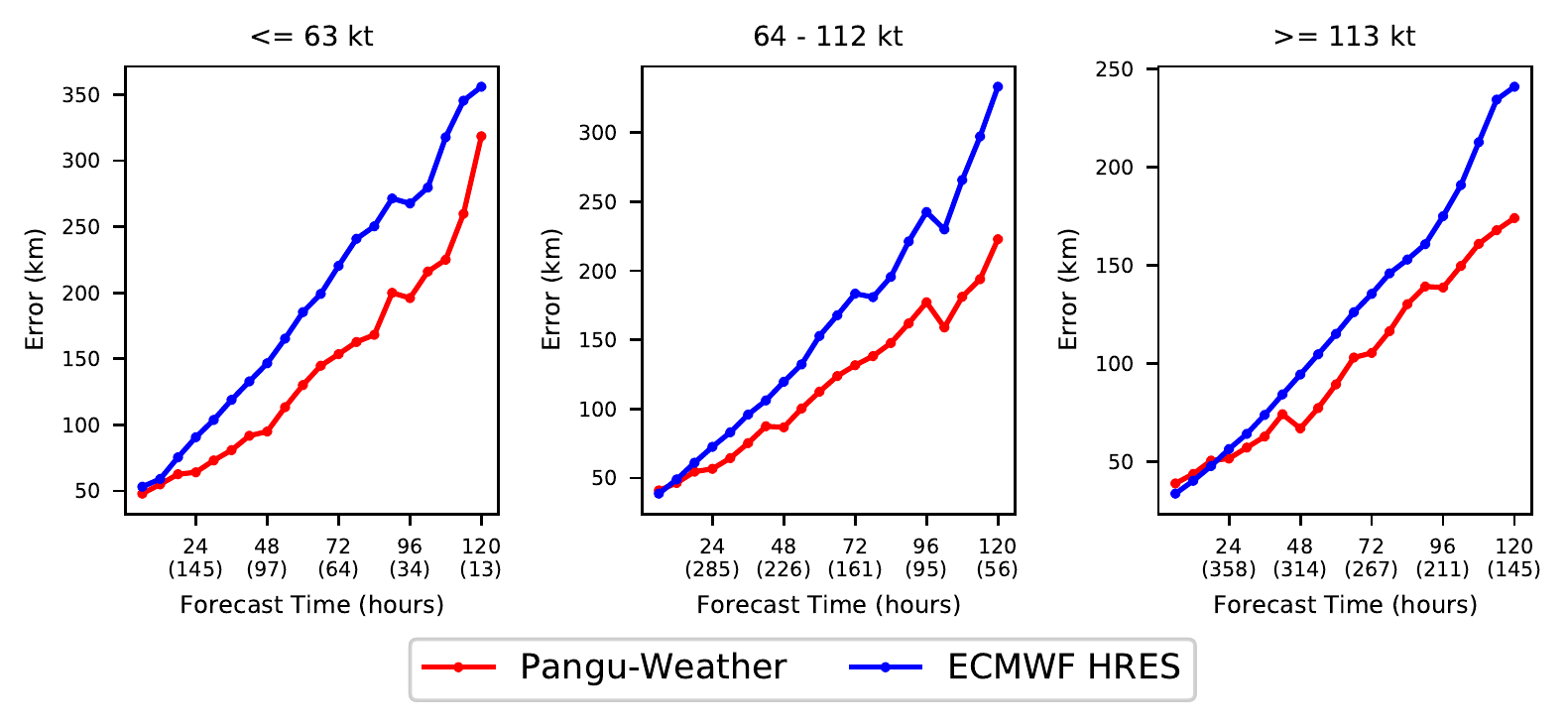}
\caption{The comparison of mean direct position errors of tropical cyclone tracking between Pangu-Weather and ECMWF-HRES, where the results are obtained by averaging $88$ named tropical cyclones in 2018. We show the overall results (top), the results with respect to different basins, and the results with respect to different intensities (bottom).}
\label{fig:cyclone_postion_error}
\end{figure}

\begin{figure*}
\centering
\begin{adjustbox}{width=\textwidth}
\begin{tikzpicture}
\node[inner sep=0pt] (tp1) at (0,0){\includegraphics[width=8.5cm,height=6.5cm]{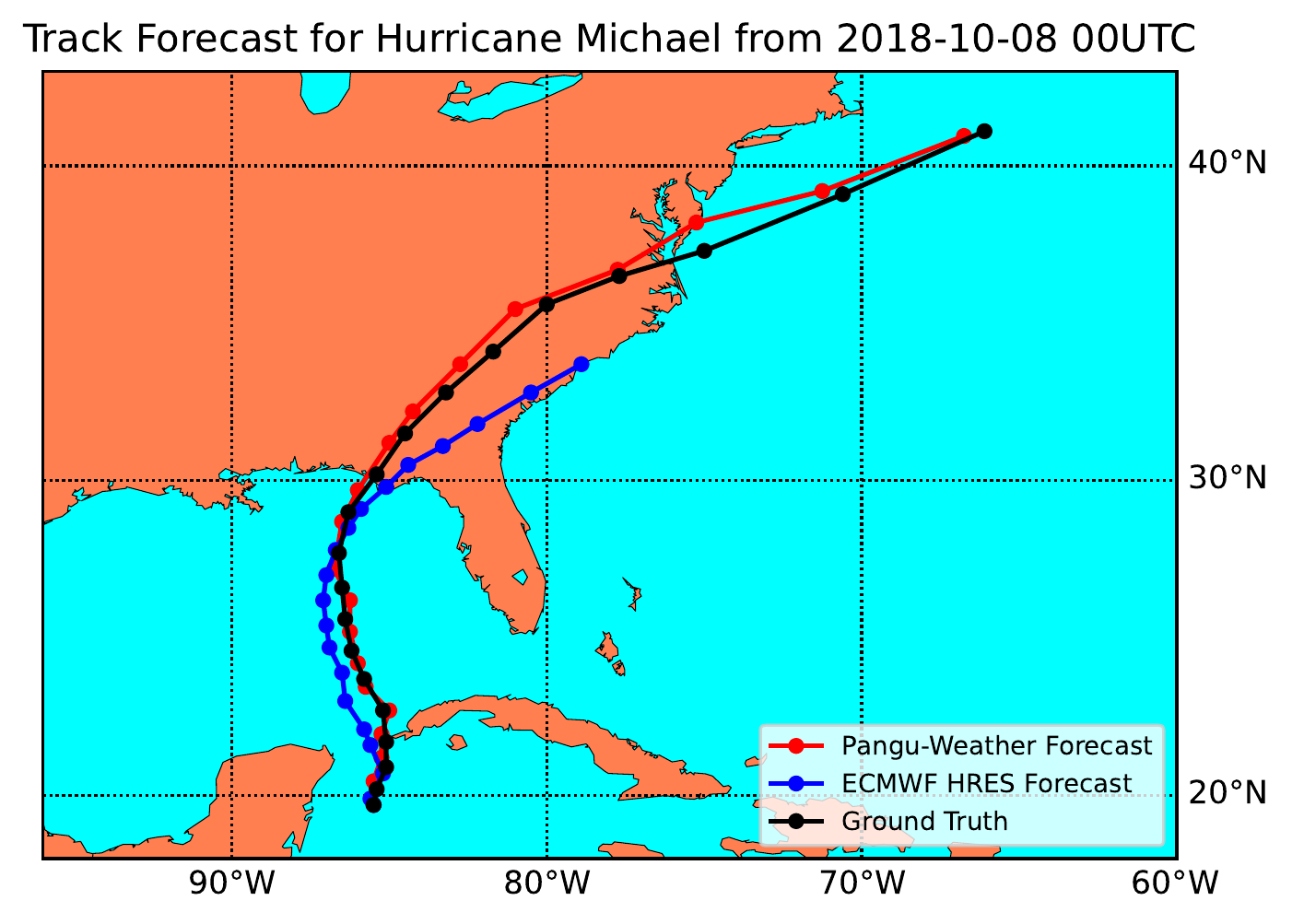}};
\draw [-stealth](-1.73,0.135) -- (4.23,2);
\draw [-stealth](-1.73,0.135) -- (4.23,-2);
\draw[draw=black] (4.25,-3.15) rectangle ++(8.5,6.3);
\node[inner sep=0pt] (tp2) at (8.5,0){\includegraphics[width=8.5cm,height=6.5cm]{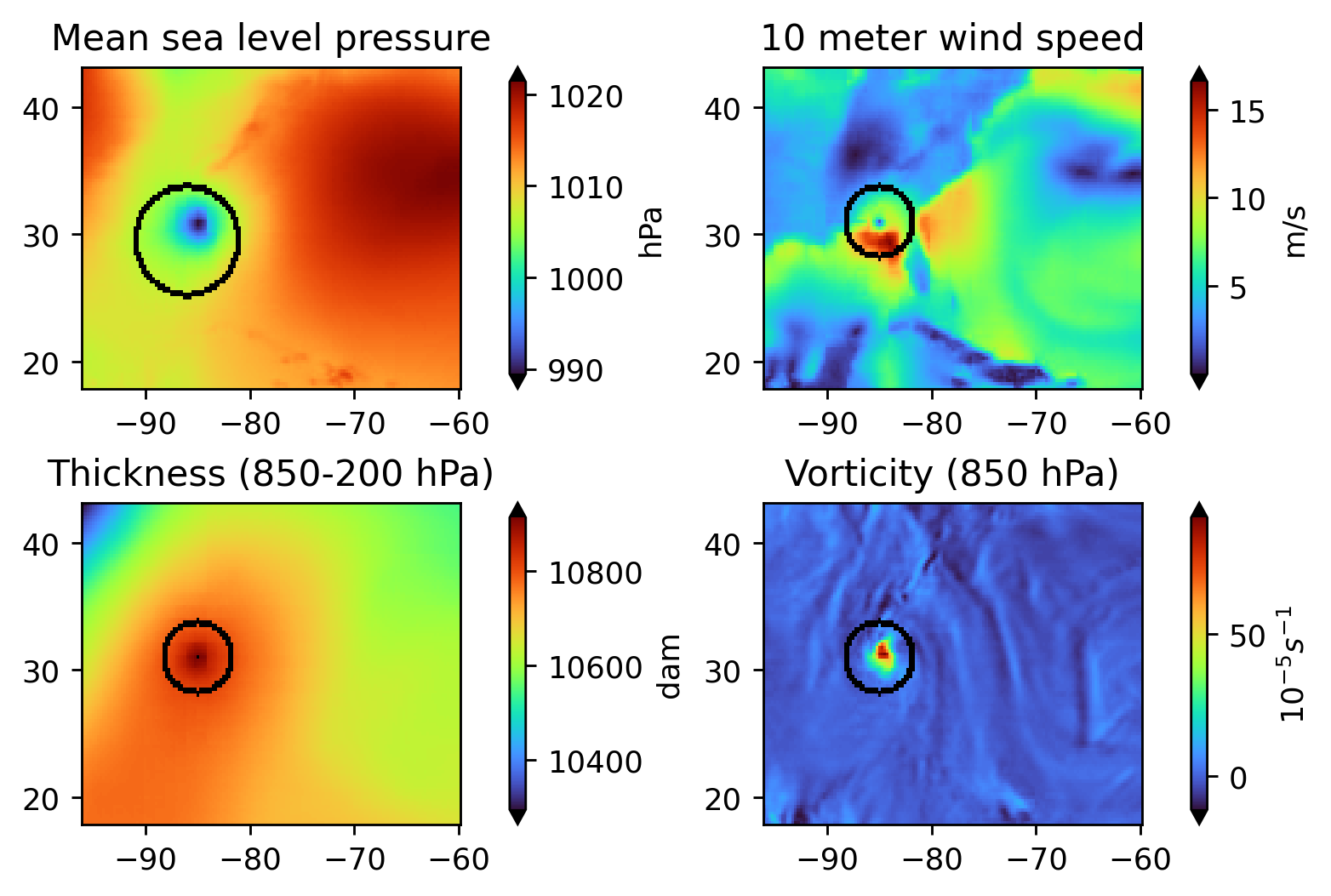}};
\node[inner sep=0pt] (vis1) at (0,-6.5){\includegraphics[width=8.5cm,height=6.5cm]{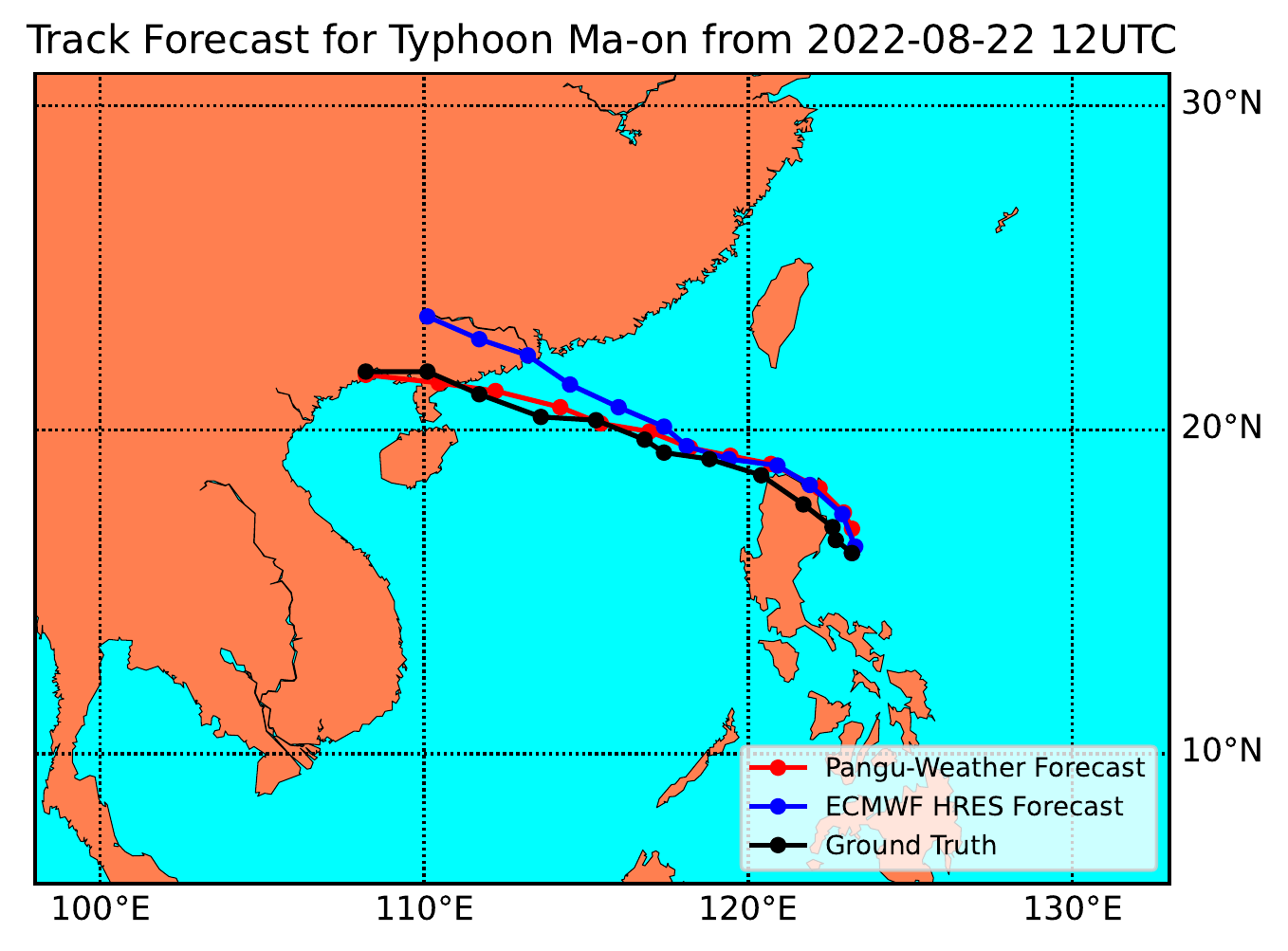}};
\draw [-stealth](-1.85,-5.83) -- (4.23,-4.5);
\draw [-stealth](-1.85,-5.83) -- (4.23,-8.5);
\draw[draw=black] (4.25,-9.65) rectangle ++(8.5,6.3);
\node[inner sep=0pt] (vis2) at (8.5,-6.5){\includegraphics[width=8.5cm,height=6.5cm]{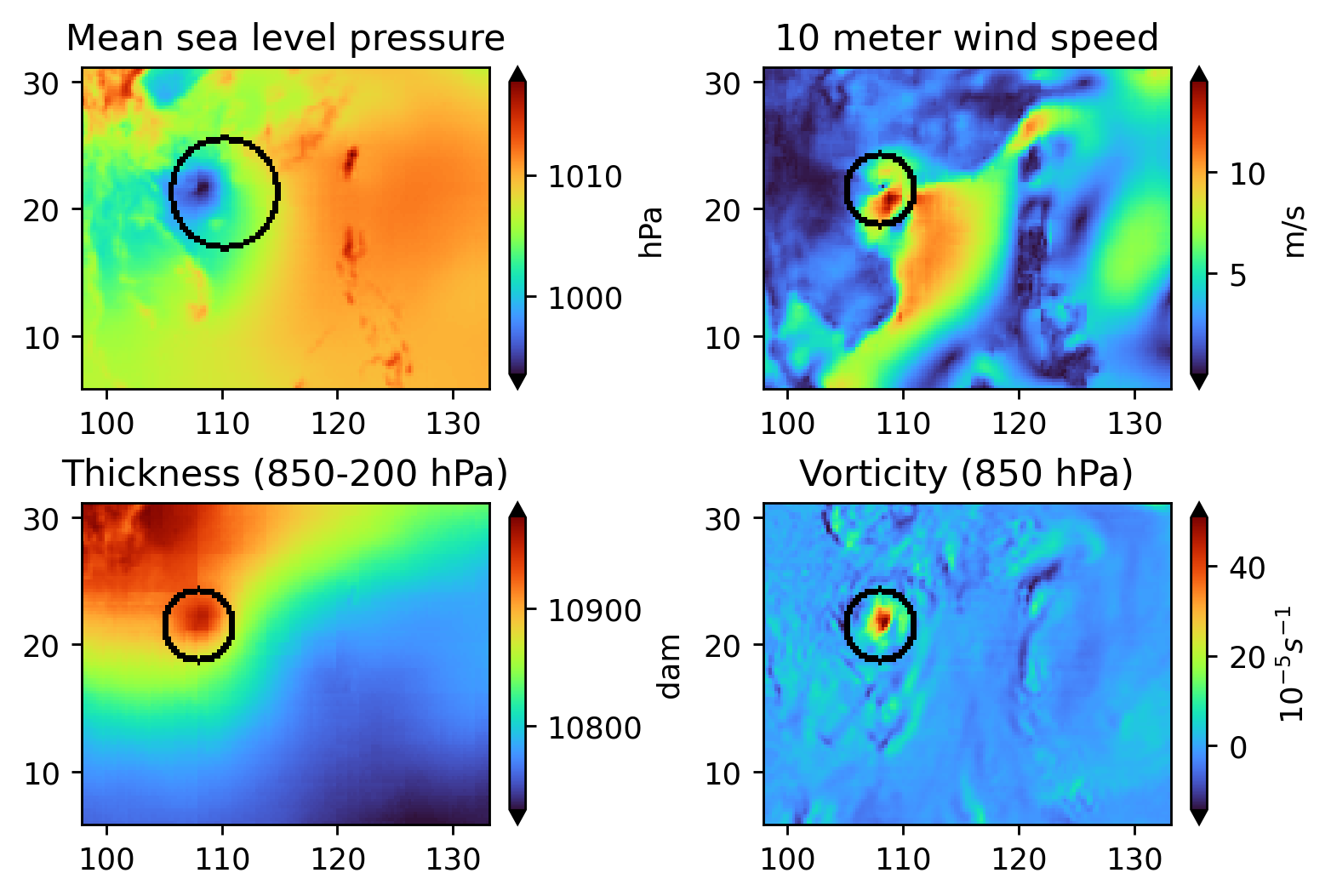}};
\end{tikzpicture}
\end{adjustbox}
\caption{\textbf{Left}: the tracking of cyclone eyes for Hurricane Michael (2018-13) and Typhoon Ma-on (2022-09) by Pangu-Weather and ECMWF-HRES, with a comparison to the ground-truth (by IBTrACS). \textbf{Right}: an illustration of the tracking process, where we use Pangu-Weather as an example. It locates cyclone eye by checking four variables (from forecast results), namely, mean sea level pressure, 10m wind speed, thickness between $850\mathrm{hPa}$ and $200\mathrm{hPa}$, and vorticity of $850\mathrm{hPa}$). The displayed figures correspond to the forecast of these variables at a forecast time of $72$ hours, and the forecast of cyclone eye is indicated using the tail of arrows.}
\label{fig:hurricane_michael}
\end{figure*}
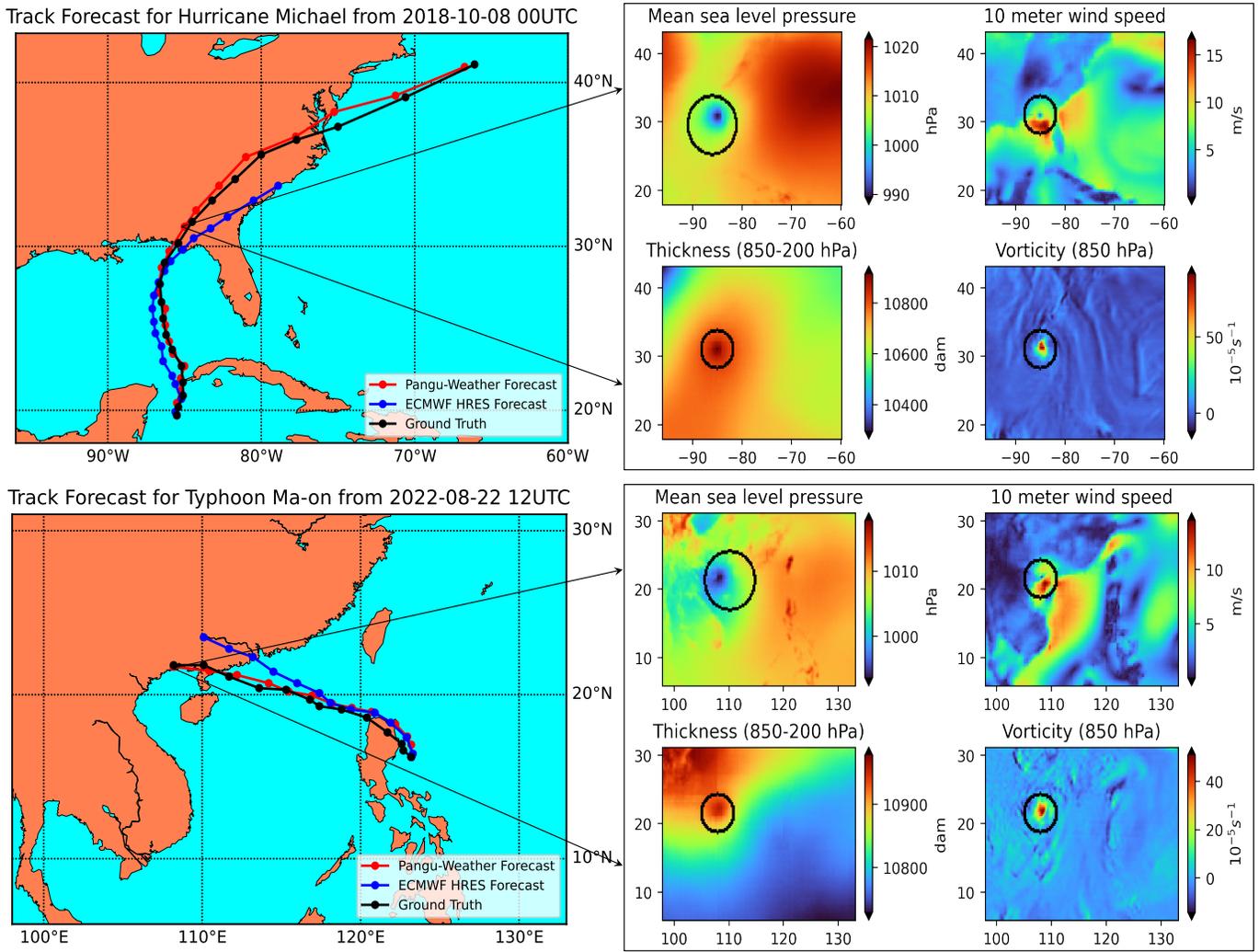

\begin{figure*}
\centering
\includegraphics[width=3cm]{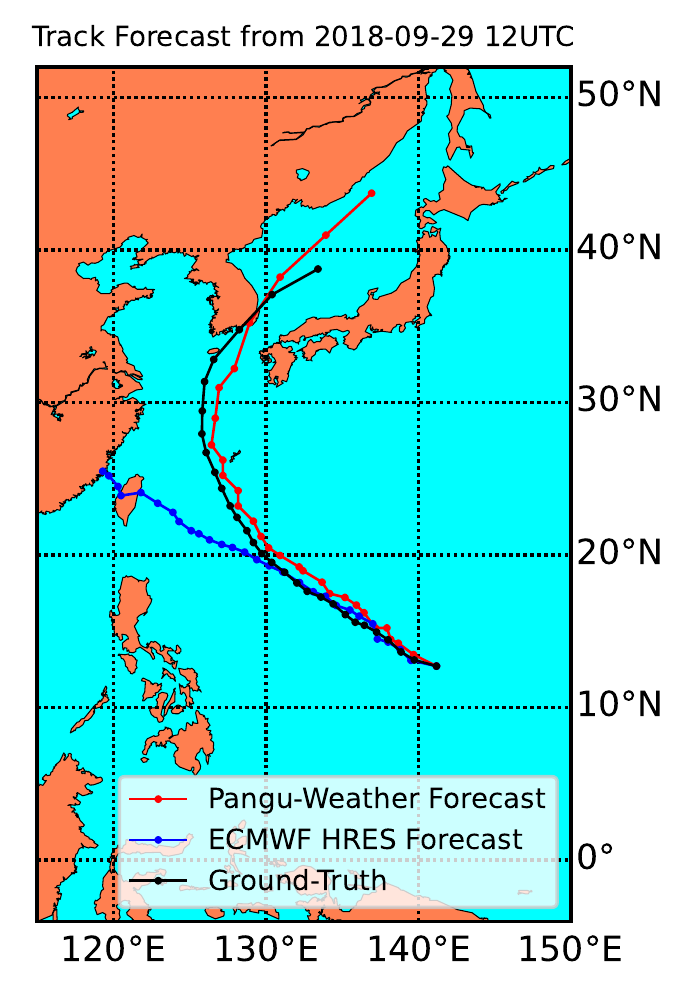}\hfill
\includegraphics[width=3cm]{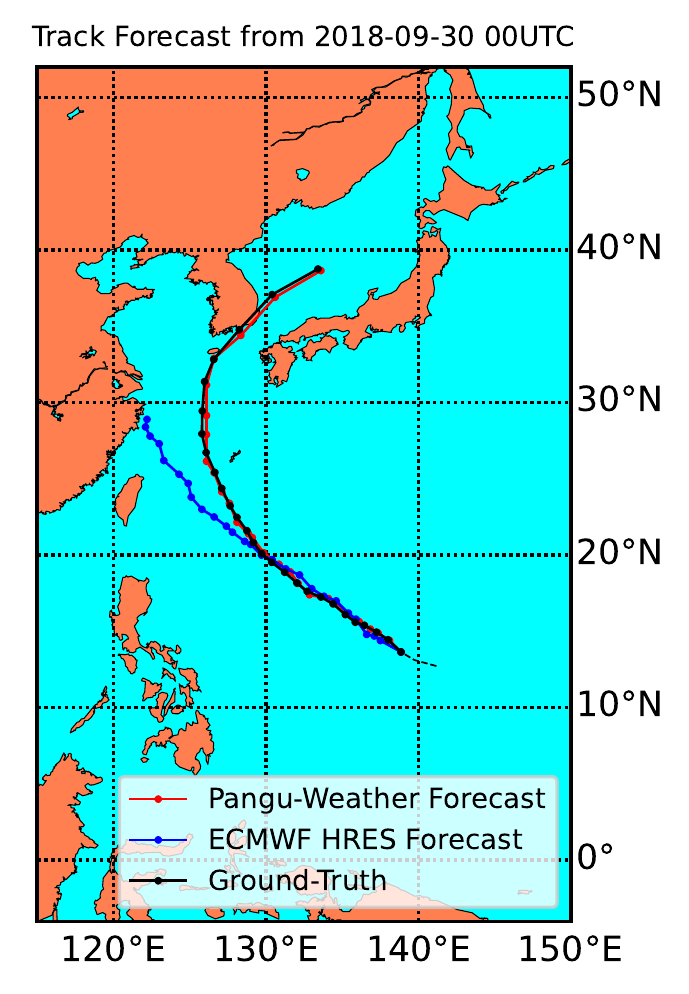}
\includegraphics[width=3cm]{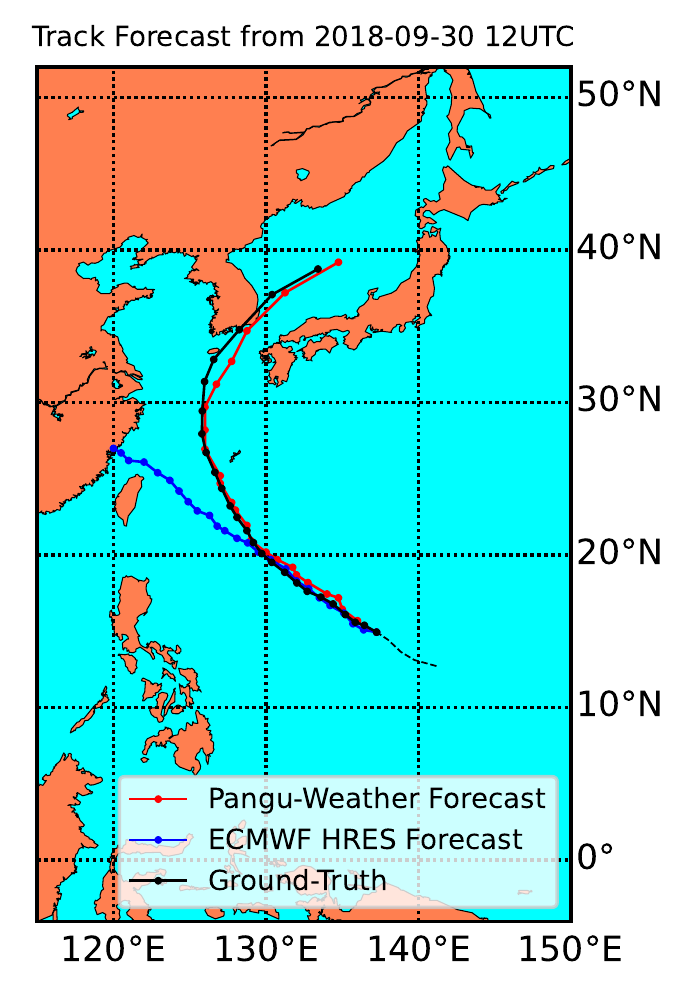}\hfill
\includegraphics[width=3cm]{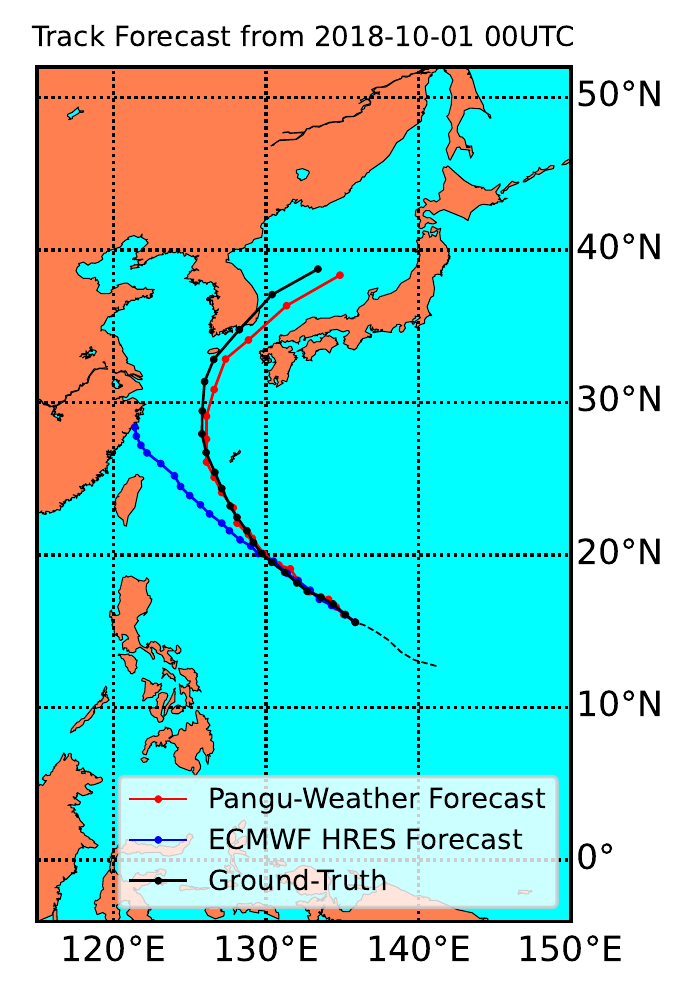}
\includegraphics[width=3cm]{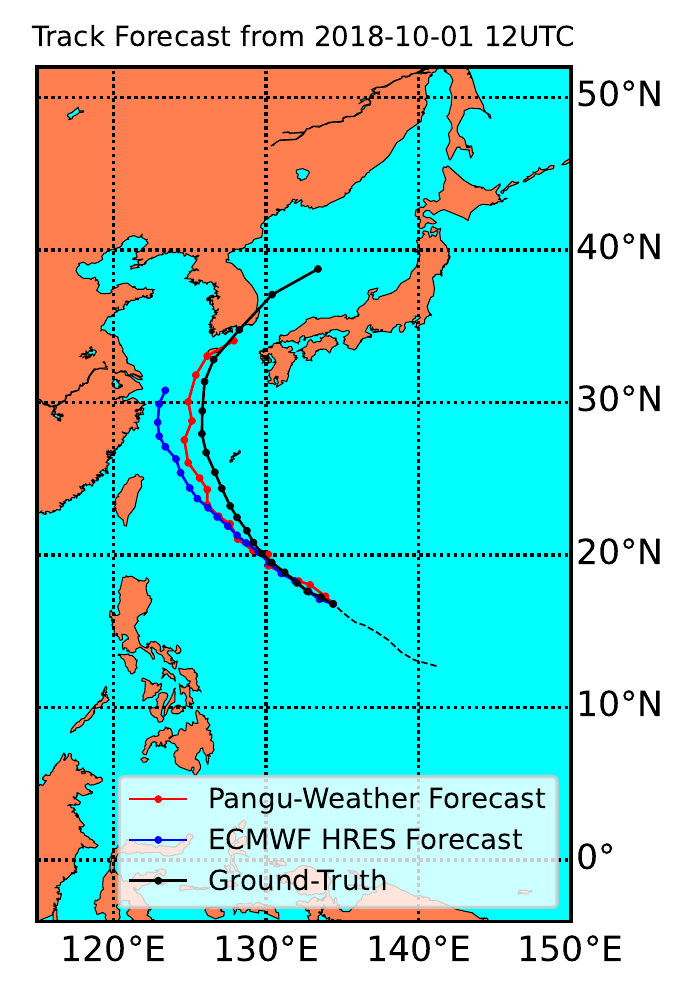}\hfill
\includegraphics[width=3cm]{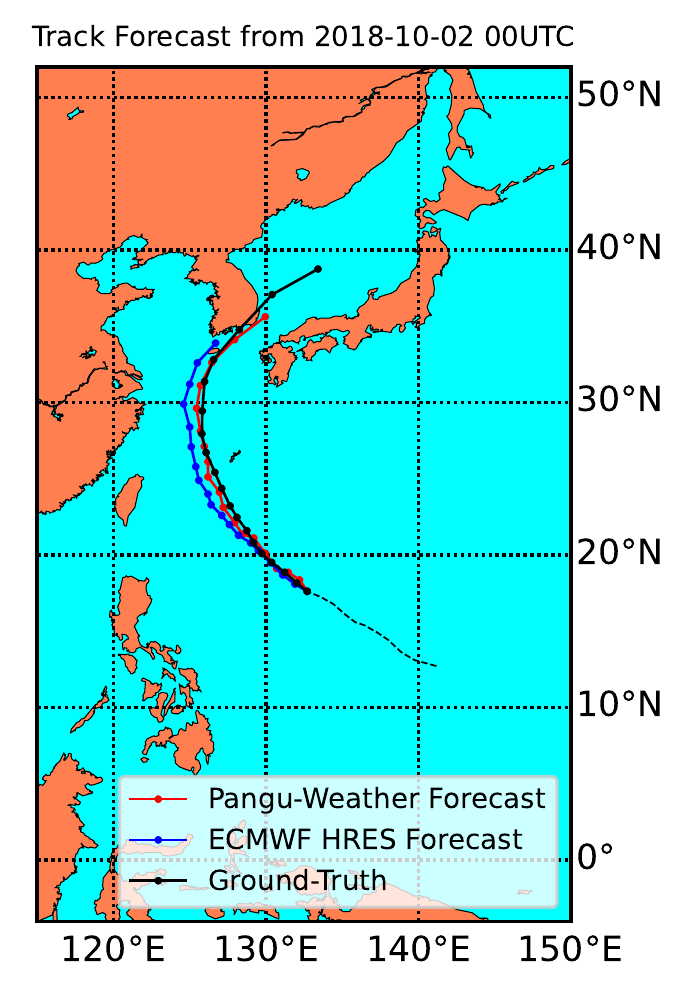}
\caption{The dynamic tracking results of cyclone eyes for Typhoon Kong-rey (2018-25) by Pangu-Weather and ECMWF-HRES, with a comparison to the ground-truth (by IBTrACS). We show six time points with the first one being 12:00 UTC, September 29th, 2018, and the time gap between neighboring points being $12$ hours. The historical (observed) path of cyclone eyes is shown in dashed. Mind the significant difference between Pangu-Weather and ECMWF-HRES (Pangu-Weather is significantly better) at the middle four time points.}
\label{fig:time_series_cyclone}
\end{figure*}

\subsubsection{Tracking Tropical Cyclones}
\label{results:extreme:cyclones}

We study a special case of extreme weather forecast, namely, tracking tropical cyclones. Note that we follow the conventions to focus on forecasting the eye of tropical cyclones rather than the intensity\footnote{Due to the limited resolution of EDA systems, reanalysis data like ERA5 always underestimate cyclones intensity (\textit{e.g.}, minimum pressure and maximum wind speed) significantly~\cite{era5_cyclone1,era5_cyclone2,ecmwf_tropical_cyclone_activity}. Trained on ERA5, it is difficult for Pangu-Weather to forecast the intensity accurately (\textit{e.g.}, the predicted minimum pressure is often $50\mathrm{hPa}$ higher than the ground-truth), while the path tracking accuracy is reasonable (see the later results). In the future, if higher-resolution, unbiased weather data (especially tropical cyclone data) are provided, it is very likely that Pangu-Weather can be directly trained or fine-tuned on these data for more accurate intensity forecast.}. Hence, in this part, we report the averaged distance between the ground-truth and predicted cyclone eyes.


To track the eye of a tropical cyclone, we follow~\cite{ECMWF_2004_track_methods} to find the local minimum of mean sea level pressure (MSLP). Specifically, we follow~\cite{ecmwf_tropical_cyclone_activity} to set the lead time to be $6$ hours. Given the starting time point and the corresponding (initial) position of a cyclone eye, we iteratively call for forecasting the $6$-hour-later weather states and look for a local minimum of MSLP that satisfies the following conditions:
\begin{itemize}
\item There is a maximum of $850\mathrm{hPa}$ relative vorticity that is larger than $5\times10^{-5}$ within a radius of $278\mathrm{km}$ for the Northern Hemisphere, or a minimum that is smaller than $5\times10^{-5}$ for the Southern Hemisphere.
\item There is a maximum of thickness between $850\mathrm{hPa}$ and $200\mathrm{hPa}$ within a radius of $278\mathrm{km}$ when the cyclone is extratropical.
\item The maximum 10m wind speed is larger than $8\mathrm{m}/\mathrm{s}$ within a radius of $278\mathrm{km}$ when the cyclone is on land.
\end{itemize}
Once the cyclone eye is located, the tracking algorithm continues to find the next position in a vicinity of $445\mathrm{km}$. The tracking algorithm terminates when no local minimum of MSLP is found to satisfy the above conditions.

We refer to the International Best Track Archive for Climate Stewardship (IBTrACS) project~\cite{cyclone_best_track_1,cyclone_best_track_2} which contains the best available estimations for tropical cyclones. We directly apply the above tracking algorithm on the deterministic forecast results of Pangu-Weather. We compare the tracking results to ECMWF-HRES, a strong competitor of cyclone tracking based on high-resolution ($9{km}\times9{km}$) operational weather forecast -- clearly, a higher-resolution forecast is more accurate in locating tropical cyclone eyes. The ECMWF-HRES forecast of cyclone eyes are directly downloaded from the TIGGE archive~\cite{tigger_data}. For a fair comparison, we choose the tropical cyclones in 2018 (the year of the above quantitative study of deterministic forecasts) that appeared in both the IBTrACS project and the ECMWF-HRES forecasts. This results in a dataset (which we call TC2018) with $88$ named tropical cyclones.

We quantitatively compare the forecast accuracy of Pangu-Weather and ECMWF-HRES in TC2018. The $3$-day and $5$-day mean direct position errors (for cyclone eyes) of Pangu-Weather are $120.29\mathrm{km}$ and $195.65\mathrm{km}$, respectively, which are significantly smaller than $162.28\mathrm{km}$ and $272.10\mathrm{km}$ reported by ECMWF-HRES. Figure~\ref{fig:cyclone_postion_error} plots the mean direct position errors with respect to forecast time. One can see a clear advantage of Pangu-Weather over ECMWF-HRES over the entire dataset and within subsets of different basins or different intensities. Inheriting the property of deterministic forecast, the advantage becomes more significant when forecast time gets larger.

The tracking results of some representative cases are shown in Figures~\ref{fig:overview} and~\ref{fig:hurricane_michael}. We study four representative cases, three in 2018 and one in 2022. For Michael and Ma-on, we set the starting time point to be the earliest one in the ECMWF-HRES forecast, while for Kong-rey and Yutu, the starting points are postponed for a few days for better visualization. We use Pangu-Weather to forecast the entire cyclone path (\textit{i.e.}, until the cyclone dissipates), and compare the tracking results to ECMWF-HRES and the ground-truth. Again, Pangu-Weather produces much more accurate tracking results compared to ECMWF-HRES, and the advantage becomes large as forecast time increases. Below, we analyze these cases one-by-one.
\begin{itemize}
\item \textbf{Typhoon Kong-rey (2018-25)}\footnote{\textsf{https://en.wikipedia.org/wiki/Typhoon\_Kong-rey\_(2018)}} is one of the most powerful tropical cyclones worldwide in 2018. It caused $4$ fatalities and \$$171.5$ million damage. As shown in Figure~\ref{fig:overview}, ECMWF-HRES forecasts that Kong-rey would land on China, but it actually did not. Pangu-Weather, instead, produces accurate tracking results which almost coincide with the ground-truth. Also, Figure~\ref{fig:time_series_cyclone} shows the tracking results of Pangu-Weather and ECMWF-HRES at different time points -- the forecast of Pangu-Weather barely changes with time, yet ECMWF-HRES arrives at the conclusion that Kong-rey would not land on China more than $48$ hours later than Pangu-Weather.
\item \textbf{Typhoon Yutu (2018-26)}\footnote{\textsf{https://en.wikipedia.org/wiki/Typhoon\_Yutu}} is an extremely powerful tropical cyclone that caused catastrophic destruction in the Mariana Islands and the Philippines. It also ties Kong-rey as the most powerful tropical cyclone worldwide in 2018, resulting in $30$ fatalities and \$$854.1$ million damage. As shown in Figure~\ref{approach:overview}, Pangu-Weather makes the correct forecast (Yutu goes to the Philippines) as early as $6$ days before landing, while the forecast of ECMWF-HRES is dramatically incorrect (Yutu makes a big turn and heads to the northeast). ECMWF-HRES forecasts the correct direction more than $48$ hours later than Pangu-Weather.
\item \textbf{Hurricane Michael (2018-13)}\footnote{\textsf{https://en.wikipedia.org/wiki/Hurricane\_Michael}} is the strongest hurricane of the 2018 Atlantic hurricane season. Michael became a Category-5 hurricane and landed on Florida on October 10th, 2018, resulting in $74$ fatalities and \$$25.5$ billion damage. As shown in Figure~\ref{fig:hurricane_michael}, with a starting time that is more than $3$ days earlier than landing, Both Pangu-Weather and ECMWF-HRES forecast the landfall on Florida, but the delay of predicted landing time is only $3$ hours for Pangu-Weather but $18$ hours for ECMWF-HRES. In addition, Pangu-Weather shows great advantages in tracking Michael after it landed, while the tracking of ECMWF-HRES is much shorter and shifts to the east obviously.
\item \textbf{Typhoon Ma-on (2022-09)}\footnote{\textsf{https://en.wikipedia.org/wiki/Tropical\_Storm\_Ma-on\_(2022)}} is a severe tropical storm that impacted the Philippines and China. Ma-on landed over Maconacon, Philippines on August 23rd and made a second landfall over Maoming, China on August 25th, resulting in $7$ fatalities and \$$9.13$ million damage. As shown in Figure~\ref{fig:hurricane_michael}, when the starting time point is about $3$ days earlier than landing, ECMWF-HRES produces a wrong forecast that Ma-on would land on Zhuhai, China, while the forecast of Pangu-Weather is about right.
\end{itemize}


\begin{figure*}
\centering
\includegraphics[width=17cm]{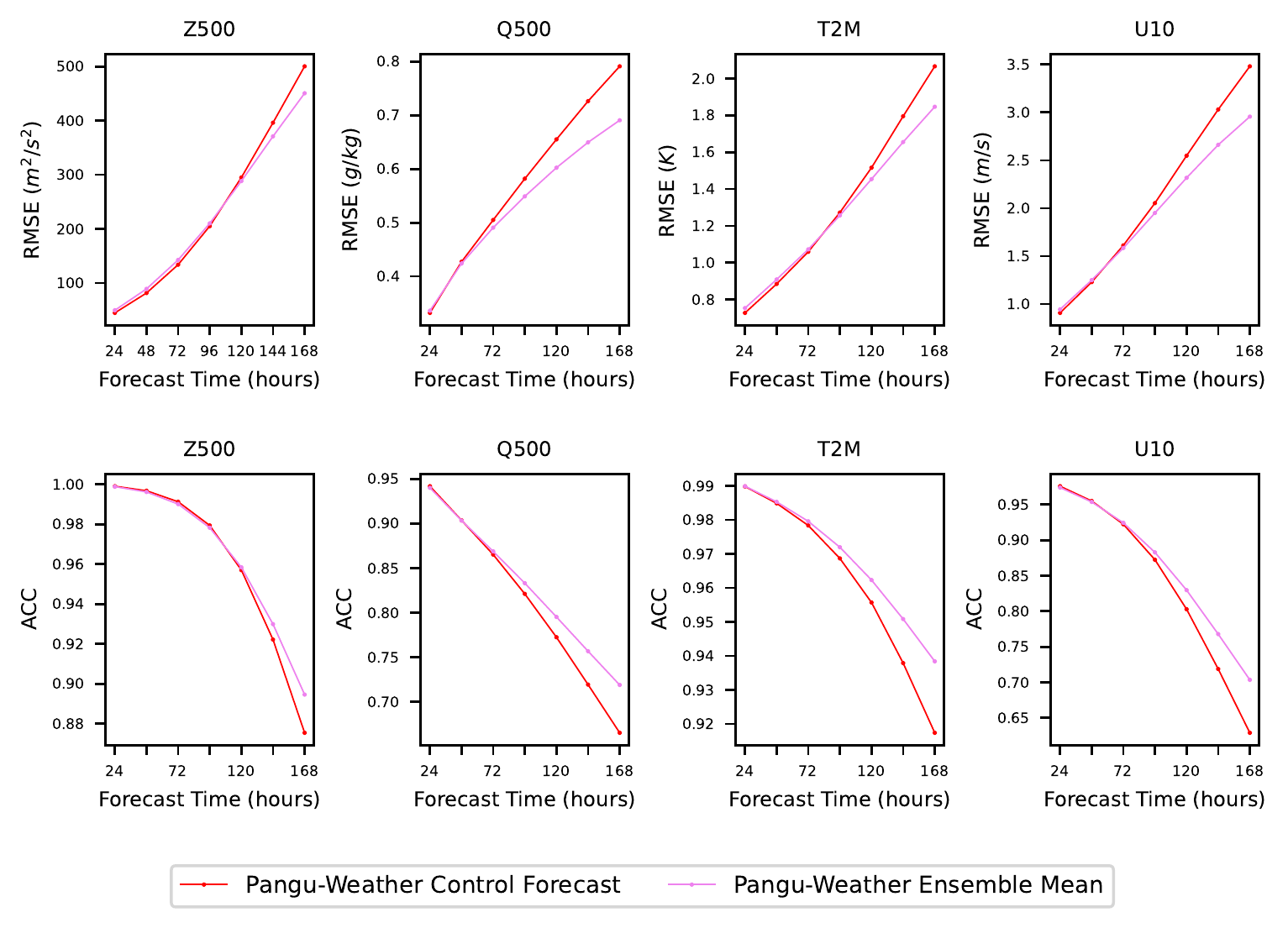}
\caption{Comparison of control forecast and ensemble forecast results of Pangu-Weather. We display the latitude-weighted RMSE (lower is better) and ACC (higher is better) for two upper-air variables ($500\mathrm{hPa}$ geopotential, Z500, and $500\mathrm{hPa}$ specific humidity, Q500) and two surface variables (2m temperature, T2M, and $u$-component of 10m wind speed, U10).}
\label{fig:ensemble_forecast}
\end{figure*}

The much better tracking results are directly owed to the higher deterministic forecast accuracy of Pangu-Weather. In the right part of Figure~\ref{fig:hurricane_michael}, we show how Pangu-Weather tracks Hurricane Michael and Typhoon Ma-on following the specified tracking algorithm. Among the four variables, mean sea level pressure and 10m wind speed are directly produced by deterministic forecast, and thickness and vorticity are mainly derived from geopotential and wind speed. This indicates that Pangu-Weather can produce intermediate results that support cyclone tracking, which further assists meteorologists in understanding and exploiting the tracking results.

In summary, the advantages of Pangu-Weather in tracking tropical cyclone eyes are mainly inherited from the good practice of deterministic forecast, in particular, the forecast of mean sea level pressure that is critical for locating cyclone eyes. Arguably, the proposed 3DEST architecture incorporates 3D information to improve the accuracy of important variables such as geopotential, wind speed, and mean sea level pressure. On the other hand, we notice that Pangu-Weather still heavily underestimates the intensity of tropical cyclones, arguably due to the same weakness of the ERA5 data. In the future, we expect higher-resolution, more accurate training data can be established for fine-tuning Pangu-Weather for these specific scenarios of extreme weather forecast. We also welcome meteorologists to offer expertise to improve the forecast of intensity.



\subsection{Ensemble Forecast}
\label{results:ensemble}

A core goal of ensemble weather forecast is to investigate the uncertainty of forecast systems, \textit{i.e.}, the change of forecast results with respect to small perturbations. Researchers mainly considered two types of uncertainty added to by (i) initial weather states and (ii) model parameters. The purpose is for diagnosing errors in observed or reanalyzed data (\textit{e.g.}, caused by assimilation~\cite{data_assimilation_review}) and bias of both NWP and AI-based models (\textit{e.g.}, biasing towards false patterns).

The methodology of ensemble forecast mainly involves adding noise to either initial weather states or model parameters and observing the change of forecast results. Either of them requires performing the inference multiple times. Pangu-Weather, as an AI-based method, enjoys a much faster inference speed than conventional NWP methods, \textit{e.g.}, more than $10\rm{,}000\times$ faster than operational IFS. This offers an opportunity of performing large-member ensemble forecast in relatively low computational costs. In what follows, we offer a preliminary study of ensemble forecast based on Pangu-Weather, yet we believe that meteorologists can offer professional knowledge to further utilize the ability of ensemble forecast.

In this paper, we mainly investigate the first line that adds perturbations to initial weather states. For simplicity, we follow FourCastNet~\cite{fourcastnet} to set the perturbations to be random Perlin noise, while we believe that richer meteorologic knowledge can assist us in developing more advanced ensemble methods (\textit{e.g.}, based on singular vectors~\cite{singular_vectors_review}). Mathematically, let the initial weather state be $\mathbf{A}_t^\ast$, and we randomly generate $S=99$ Perlin noise vectors of the same size of $\mathbf{A}_t^\ast$, denoted as $\mathbf{P}_1,\ldots,\mathbf{P}_S$. The initial states are perturbed into $\mathbf{A}_t^\ast+\eta\mathbf{P}_s$, where $\eta=0.2$ is the coefficient that controls the noise amplitude, and $s=0,1,\ldots,S$ where $s=0$ implies that no noise is added, \textit{i.e.}, $\mathbf{P}_0\equiv\mathbf{0}$. We feed all the $S+1$ initial states to the trained model and average the outputs as the final ensemble forecast result. Experiments are still performed on the ERA5 weather data in 2018, where deterministic forecast is taken as a natural baseline. As shown in Figure~\ref{fig:ensemble_forecast}, the accuracy of $100$-member ensemble forecast is slightly worse than single-member deterministic forecast in short-range (\textit{e.g.}, $1$-day) weather forecast, but is significantly higher than deterministic forecast when forecast time is longer than $5$ days. This aligns with our intuition and the observations of prior work~\cite{fourcastnet}, indicating that large-member ensemble forecast is especially useful when single-model accuracy becomes lower, yet it risks introducing unexpected noise that may cause accuracy drop when the deterministic forecast is accurate enough. In addition, ensemble forecast brings more benefits to the non-smooth variables such as $500\mathrm{hPa}$ specific humidity (Q500) and 10m surface wind speed (U10), \textit{e.g.}, the latitude-weighted RMSEs of $7$-day forecast for Z500 and U10 are reduced from $500.3$ and $3.48$ to $450.6$ and $2.96$, with relative drops of $10\%$ and $15\%$, respectively.

Temporarily, we do not study another line (\textit{i.e.}, adding perturbations to model parameters) because Pangu-Weather is based on deep neural networks that contains hundreds of millions of parameters, unlike conventional NWP models~\cite{model_uncertainty_review, model_error_ecmwf_early} that contain much fewer yet physically meaningful parameters. In addition, the learned parameters are highly sensitive to a few random factors during the training procedure (\textit{e.g.}, random seeds, data sampling strategies, \textit{etc.}) and thus are difficult to be perturbed for specific purposes. In the future, with the guidance from meteorologists, we expect to fine-tune the base Pangu-Weather models into a series of `child models' for manipulating different factors.

\section{Conclusions and Future Remarks}
\label{conclusions}

In this paper, we present Pangu-Weather, an AI-based system for numerical weather forecast. The technical contribution involves (i) designing the 3D Earth-specific transformer (3DEST) architecture and (ii) applying the hierarchical temporal aggregation strategy. By training deep neural networks on $39$ years of global weather data, Pangu-Weather, for the first time, surpasses the conventional NWP methods in terms of both accuracy and speed. Being efficient in inference, Pangu-Weather opens a window for meteorologists to integrate their knowledge to AI-based methods for more exciting applications.

Looking into the future, we expect that computational resource is the key to further improving the accuracy of weather forecast. According to our experiments, the training procedure has not yet arrived at full convergence, and there is much room left in terms of (i) incorporating more observation factors, (ii) integrating the time dimension and training 4D deep networks, and (iii) simply using deeper and/or wider networks. All of these require more powerful GPUs with larger memory and higher FLOPs.

\section*{Acknowledgments}

We would like to thank ECMWF for offering the ERA5 dataset and the TIGGE archive. Without such selfless dedication, this research would never become possible. We thank NOAA National Centers for Environmental Information for the IBTrACS dataset. We thank other members of the Pangu team for instructive discussions and support in computational resource. Our appreciation also goes to the Integration Verification team of Huawei Cloud EI that offers us a platform of high-performance parallel computing.

\ifCLASSOPTIONcaptionsoff
  \newpage
\fi



%
\bibliographystyle{IEEEtran}
\bibliography{IEEEabrv,refs}

%

\end{document}